\documentclass[journal]{IEEEtran}

\ifCLASSINFOpdf
  \usepackage[pdftex]{graphicx}
  \graphicspath{{./}}
  \DeclareGraphicsExtensions{.pdf}
\else
  \usepackage[dvips]{graphicx}
  \graphicspath{{./}}
  \DeclareGraphicsExtensions{.pdf}
\fi

\usepackage[cmex10]{amsmath}

\usepackage[caption=false,font=footnotesize]{subfig}

\usepackage{multirow}
\usepackage{dsfont}
\usepackage{amssymb}
\usepackage{calc}

\newcommand{\D}{\mathbf{D}}

\newcommand{\I}{\mathbf{I}}
\newcommand{\Ha}{\mathbf{H}}
\newcommand{\B}{\mathbf{B}}

\newcommand{\A}{\mathbf{A}}

\newcommand{\F}{\mathbf{F}}

\newcommand{\R}{\mathbf{R}}

\newcommand{\f}{\mathbf{f}}
\newcommand{\g}{\mathbf{g}}
\newcommand{\x}{\mathbf{x}}
\newcommand{\y}{\mathbf{y}}

\newcommand{\n}{\mathbf{n}}
\newcommand{\bb}{\mathbf{b}}

\newcommand{\rr}{\mathbf{r}}
\newcommand{\cc}{\mathbf{c}}

\newtheorem{theorem}{Theorem}
\newenvironment{definition}[1][Definition]{\begin{trivlist}
\item[\hskip \labelsep {\bfseries #1}]}{\end{trivlist}}

\begin{document}

\title{Coded Acquisition of High Frame Rate Video}

\author{Reza~Pournaghi
        and~Xiaolin~Wu,~\IEEEmembership{Fellow,~IEEE}
\thanks{X.\ Wu is with the Department
of Electrical and Computer Engineering, McMaster University, Hamilton,
ON, L8S~4K1 Canada e-mail: xwu@mcmaster.ca.}
\thanks{R.\ Pournaghi is PHD student at the
Department of Electrical and Computer Engineering, McMaster University, Hamilton,
ON, L8S~4K1 Canada e-mail: pournaghi@grads.ece.mcmaster.ca.}}


\maketitle

\begin{abstract}
\label{sec:abstrc}
High frame video (HFV) is an important investigational tool in sciences, engineering and military.  In ultra-high speed imaging, the
obtainable temporal, spatial and spectral resolutions are limited by the sustainable throughput of in-camera mass memory, the lower bound of exposure time, and illumination conditions. In order to break these bottlenecks, we propose a new coded video acquisition framework that employs $K \geq 2$ conventional cameras, each of which makes random measurements of the 3D video signal in both temporal and spatial domains.  For each of the $K$ cameras, this multi-camera strategy greatly relaxes the stringent requirements in memory speed, shutter speed, and illumination strength.  The recovery of HFV from these random measurements
is posed and solved as a large scale $\ell_1$ minimization problem by exploiting joint temporal and spatial sparsities of the 3D signal.  Three coded video acquisition techniques of varied trade offs between performance and hardware complexity are developed: frame-wise coded acquisition, pixel-wise coded acquisition, and column-row-wise coded acquisition.  The performances of these techniques are analyzed in relation to the sparsity of the underlying video signal.  Simulations of these new HFV capture techniques are carried out and experimental results are reported.
\end{abstract}

\begin{IEEEkeywords}
High frame rate video, coded acquisition, random sampling, sparse representations, digital cameras.
\end{IEEEkeywords}

%
\IEEEpeerreviewmaketitle

\section{Introduction}
\label{sec:intro}

\IEEEPARstart{H}{igh} frame rate video (HFV) enables investigations of high speed physical phenomena like explosions, collisions, animal kinesiology, and etc.  HFV cameras find many applications in sciences, engineering research, safety studies, entertainment and defense \cite{WilburnJoshiVaish04}.  Compared with conventional (low-speed) cameras, HFV (high-speed) cameras are very expensive.  Despite their high costs,
HFV cameras are still limited in obtainable joint temporal-spatial resolution, because current fast mass data storage devices (e.g., SSD) do not have high enough write speed to continuously record HFV at high spatial resolution.  In other words, HFV cameras have to compromise spatial resolution in quest for high frame rate.  For instance, the HFV camera Phantom v710 of Vision Research can offer a spatial resolution of $1280 \times 800$ at $7530$ frame per second (fps), but it has to reduce the spatial resolution to $128\times128$ when operating at $215600$ fps.  This trade-off between frame rate and spatial resolution is forced upon by the mismatch between ultra-high data rate of HFV and limited bandwidth of in-camera memory.
Also, application scenarios exist when the raw shutter speed is restricted by low illumination of the scene. Needless to say, these problems are aggravated if high spectral resolution
of HFV is also desired.
No matter how sophisticated sensor and memory technologies become, new, more exciting and exotic applications will always present themselves that require imaging of
ever more minuscule and subtle details of object dynamics.  It is, therefore, worthy and satisfying to research on camera systems and accompanying image/video processing techniques
that can push spatial, temporal and spectral resolutions to the new limit.

One way of breaking the bottlenecks for extreme imaging of high temporal-spatial-spectral fidelity is to use multiple cameras.  In this paper we propose a novel multi-camera coded video acquisition system that can capture a video at very high frame rate without sacrificing spatial resolution.  In the proposed system, $K$ component cameras are employed to collectively shoot a video of the same scene, but each camera adopts a different digitally modulated exposure pattern called coded exposure or strobing in the literature.
Coded exposure is a technique of computational photography which performs multiple exposures of the sensors in random during a frame time \cite{wandell1999multiple, pointgrey2006}.  Raskar {\em et al.} appeared to be the first to use the coded exposure technique to remove motion blurs of a single camera \cite{RaskarAgrawalTumblin06}.  Theobalt {\em et al.} used a coded exposure camera to record fast moving objects
in distinct non-overlapping positions in an image \cite{theobalt2004pitching}.

\begin{figure}[!t]
\centering
\includegraphics[width=3.0in]{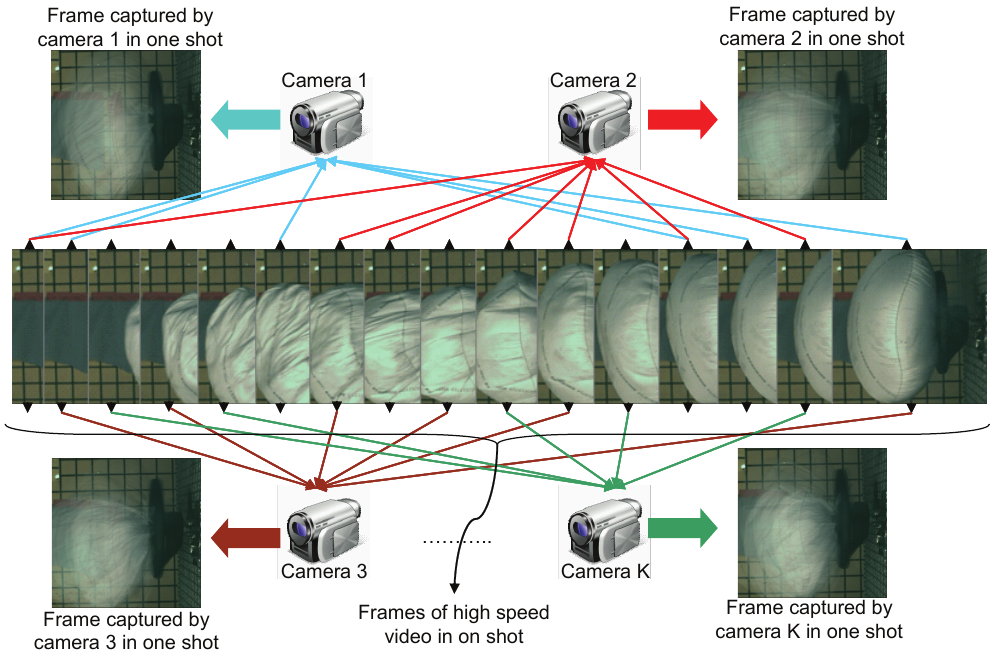}
\caption{The process of one shot by multiple coded exposure cameras.  Each camera is exposed to randomly selected high speed video frames (shown with the arrows going from the frame sequence to each camera) during a shot. The captured image is therefore blurred and contains the information of randomly selected frames.}
\label{fig:multiplcameras}
\end{figure}

In our design of HFV acquisition by multiple coded exposures,
the sequence of target HFV frames is partitioned into groups of $T$ consecutive target frames each.  Each of the $K$ component cameras of the system is modulated by a random binary sequence of length $T$ to open and close its shutter, and meanwhile the pixel sensors cumulate charges.  The camera only reads out cumulated sensor values once per $T$ target frames.  In net effect, this coded acquisition strategy reduces the memory bandwidth requirement of all $K$ cameras by $T$ folds.  Every $T$ target frames are mapped by each of the $K$ cameras to a different blurred image that is the result of summing some randomly selected target (sharp) HFV frames.
These $K$ blurred images are used to recover the corresponding $T$ consecutive HFV frames by exploiting both spatial and temporal sparsities of the HFV video signal and by solving a large-scale $\ell_1$ minimization problem.
The architecture of the coded HFV acquisition system is depicted in Fig.~\ref{fig:multiplcameras}.

The proposed multi-camera coded video acquisition approach is a way of
randomly sampling the 3D video signal independent of signal structures.  The objective is to recover an HFV signal from a far fewer number of measurements than the total number of pixels of the video sequence.  In this spirit of reduced sampling, the proposed coded HFV acquisition approach is similar to compressive sensing (CS) \cite{candss2006robust,donoho2006compressed}.  However, our research is motivated not because the spatial resolution of the camera cannot be made sufficiently high, as assumed by CS researchers in their promotion of
the "single-pixel camera" concept, rather because no existing mass storage device is fast enough to accommodate the huge data throughput of high-resolution HFV.
Very recently, Veeraraghavan {\em et al.} proposed an interesting technique to capture periodic videos by a single camera of coded exposure
\cite{veeraraghavan2010coded}.  Their success relies on the strong sparsity afforded by the signal periodicity.  But in this work, we are interested in general-purpose high-speed photography, for which, we believe, a multi-camera approach is necessary.

The idea of using multiple cameras and/or coded exposure technique to image high-speed phenomena was pioneered by Muybridge.  He used multiple triggered cameras to capture high speed motion of animals \cite{muybridge1907animals}.  Ben-Ezra and Nayar combined a high spatial resolution still camera and a high-speed but low resolution video camera to achieve high temporal and spatial resolution \cite{ben2004motion}.
Wilburn {\em et al.} proposed a multicamera HFV acquisition system \cite{WilburnJoshiVaish04}.
They used a dense array of $K$ cameras of frame rate $r$ to capture high speed videos of frame rate $h= r K$.  The idea is to stagger the start times of the exposure durations of these $K$ cameras
by $1/h$.  The captured frames are then interleaved in chronological order to generate HFV \cite{WilburnJoshiVaish04}.
Shechtman {\em et al.} fused information from multiple low resolution video sequences of the same scene to construct a video sequence of high space-time resolution using supper-resolution techniques \cite{ShechtmanCaspiIrani02}.

For different trade-offs between sampling efficiency and the hardware complexity of the multi-camera system, few different coded HFV acquisition techniques are investigated in this work.  The simplest one is called frame-wise coded video acquisition, which we outlined earlier in the introduction.  The other is pixel-wise coded video acquisition, in which each pixel is modulated by a different binary random sequence in time during the exposure, in contrast to that all pixels share the same binary modulation pattern as in frame-wise coded acquisition.  Pixel-wise coded acquisition is superior to frame-wise coded acquisition in reconstruction quality given the number of measurements.  But the former requires considerably more complex and expensive control circuitry embedded in the pixel sensor array than the latter.  To make hardware complexity manageable, we finally propose a column-row-wise coded acquisition strategy that can match the performance of pixel-wise coded acquisition but at a fraction of the cost.


The multi-camera HFV acquisition system has many advantages over existing HFV cameras. First, inexpensive conventional cameras can be used to capture high-speed videos.  This makes HFV camera systems more cost effective considering that the price differential between a 2000 fps camera and a 30 fps camera of comparable spatial resolution can be as large as one thousand holds.
Second, the new system does not compromise the spatial resolution of HFV, thanks to drastic shortening of sensor read-out time by coded acquisition.
Third, the proposed camera system architecture is highly scalable: more cameras (possibly high-speed ones) can be employed to achieve extremely high frame rate, which is otherwise unobtainable.  This is very much analogous to the multiprocessor technology that improves the computation throughput when the speed of a single CPU cannot go any higher.  Forth, the multi-camera coded acquisition system can capture HFV in low illumination conditions, which is a beneficial side effect of using low-speed cameras.

%
%

The remainder of the paper is structured as follows.
In the next section, we introduce the concept of coded exposure, and propose the HFV acquisition system using multiple coded exposure cameras.
In section \ref{sec:pixel-col-row-wis} we propose two other HFV sampling techniques called pixel-wise and column-row-wise coded exposure, as opposed to the one in section \ref{sec:MCE}.
In section \ref{sec:HFV_Recovery}, we formulate the recovery of the HFV signal in the context of sparse analysis model that exploits the temporal and spatial sparsities of the signal.
Section \ref{sec:system_attributes} reveals an important side benefit of proposed sparsity-based HFV recovery algorithm and explains how it can be used to add a degree of randomness to the measurements in spatial deomain.
Simulation results are reported in section \ref{sec:sim_res} and we conclude in section \ref{sec:conclude}.



\section{Multiple Coded Exposures}
\label{sec:MCE}
In $K$-camera coded acquisition of HFV, camera $k$, $1 \leq k \leq K$, opens and closes its shutter according to a binary pseudo random sequence $\bb_k=(b_{k,1}, b_{k,2}, \cdots, b_{k,T})$.
Through the above coded exposure process, camera $k$ produces a coded frame $\I_k$ out of every $T$ target frames $\f_1, \f_2, \cdots, \f_T$.  The coded (blended, blurred) image $\I_k$ is a function of the corresponding $T$ target (sharp) frames $\f_t$'s
\begin{equation}
\I_k = \sum_{t=1}^T b_{k,t} \f_t .
\end{equation}


Let vector $\f_{u,v}=(f_1(u,v), f_2(u,v), \cdots, f_T(u,v))'$ be the time series of pixels at spatial location $(u,v)$.  Camera $k$ modulates the time signal $\f_{u,v}$ with $\bb_k$, and makes a random measurement of $\f_{u,v}$
\begin{equation}
y_k(u,v) = \langle \bb_k, \f_{u,v} \rangle + n_k ,
\label{eq:framelevel}
\end{equation}
where $n_k$ is the measurement noise of camera $k$.
Let vector
\begin{equation}
\y_{u,v} = (y_1(u,v), y_2(u,v), \cdots, y_K(u,v))'
\end{equation}
be the $K$ random measurements of $\f_{u,v}$ made by the $K$ coded exposure cameras.
The measurement vector $\y_{u,v}$ of $\f_{u,v}$ is
\begin{equation}
\y_{u,v} = \B \f_{u,v} + \n
\end{equation}
where $\B$ is the $K\times T$ binary measurement matrix made of $K$ binary pseudo random sequences, i.e., row $k$ of matrix $\B$ is the coded exposure control sequence $\bb_k$ for camera $k$, and $\n$ is the measurement error vector.

Let the width and height of the video frame be $N_x$ and $N_y$.  For denotation convenience the three dimensional $T \times N_x \times N_y$ pixel grid of $T$ target frames is written as a super vector
$\f$, formed by stacking all the $T \!\cdot\! N_x \!\cdot\! N_y$ pixels in question.  In our design,
$K$ synchronized cameras of coded exposure are used to make $K \!\cdot\! N_x \!\cdot\! N_y$ measurements of $\f$, and let $\y$ be the vector formed by stacking all $N_x \!\cdot\! N_y$ $K$-dimensional measurement vectors $\y_{u,v}$, $1 \leq u \leq N_x$ and $1 \leq v \leq N_y$.  Then, we have
\begin{equation}
\y = \A \f + \n
\label{eq:underdet}
\end{equation}
where $\A$ is the $K \!\cdot\! N_x \!\cdot\! N_y \times T \!\cdot\! N_x \!\cdot\! N_y$ matrix of $K$ coded exposures that is made of matrix $\B$
\begin{equation}
\mathbf{A} = \left[\begin{array}{cccc}
\mathbf{B}_{K\times T} & \mathbf{0}_{K\times T} & \ldots & \mathbf{0}_{K\times T}  \\
\mathbf{0}_{K\times T} & \mathbf{B}_{K\times T} & \ldots & \mathbf{0}_{K\times T}  \\
\vdots            & \vdots            &        & \vdots  \\
\mathbf{0}_{K\times T} & \mathbf{0}_{K\times T} & \ldots & \mathbf{B}_{K\times T}
\end{array}\right]
\label{eq:matrix}
\end{equation}

The recovery of the original 3D signal $\f$ (the $T$ target HFV frames) from the $K$ images of coded exposure is, of course, a severely underdetermined inverse problem and has an infinite number of solutions.
Without any additional information about the signal or additional constraint on the measurement process, there is no hope of recovering $\f$ from (\ref{eq:underdet}).
A prior knowledge we have about the HFV signal $\f$ is the sparsity or compressibility of it in the sense that the signal has a small number of large coefficients and many zero or near zero coefficients when represented in an appropriate basis.
As the frame rate is very high, high sample correlation exists in the spatio-temporal domain, hence
the 3-dimensional Fourier transform $\Psi$ of $\f$ will make most of the entries of $\x = \Psi \f$ zero or near zero.  The sparse coefficient sequence $\x$ can be estimated from $\y$ by solving the following $\ell_1$ minimization problem:
\begin{equation}
\hat{\x} = \arg\min_{\x} \|\x\|_1 \;\;\;\; s.t. \;\;\;\; \|\A \Psi^*\x - \y\|_2 \leq \sigma
\label{eq:min:sparse:x}
\end{equation}
where $\Psi^*$ is the conjugate transpose of $\Psi$.
Once an estimate $\hat{\x}$ is obtained the HFV signal can be recovered as $\hat{\f} = \Psi^*\hat{\x}$.

When the orthobasis $\Psi$ is equal to canonical basis, i.e., the signal itself is sparse, we have $\Psi = \mathbf{I}$ (identity matrix) and $\x = \f$. Therefore the optimization problem in (\ref{eq:min:sparse:x}) changes to
\begin{equation}
\hat{\x} = \arg\min_{\x} \|\x\|_1 \;\;\;\; s.t. \;\;\;\; \|\A \x - \y\|_2 \leq \sigma .
\label{eq:min:sparse:f}
\end{equation}
In such cases the sparse signal $\x$ can be recovered from (\ref{eq:min:sparse:f}) if random measurement matrix $\A$ satisfies the following so-called
\emph{Restricted Isometry Property} (RIP) \cite{candes2008Intro}:
\begin{definition}
Matrix $\A$ satisfies RIP of order $S$ with isometry constant $\delta_S$ if
\[
(1-\delta_S)\|\x\|_2^2 \leq \|\A\x\|_2^2 \leq (1+\delta_S) \|\x\|_2^2
\]
holds for all $\x$ with $\|\x\|_0 \leq S$ where $\|\cdot\|_0$ counts the number of nonzero entries of a vector.
\end{definition}

In many applications, including HFV signal acquisition system proposed here, the signal is sparse in an orthobasis $\Psi$ other than canonical basis. For such cases the following \emph{$\Psi$-RIP} notation defined in \cite{eftekhari2012ris} is more convenient:
\begin{definition}
Let $\Psi$ denote an orthobasis in which signal $\f$ is sparse. Matrix $\A$ satisfies RIP of order $S$ in the basis $\Psi$ with isometry constant $\delta_S = \delta_S(\Psi)$ if
\[
(1-\delta_S)\|\f\|_2^2 \leq \|\A\f\|_2^2 \leq (1+\delta_S) \|\f\|_2^2
\]
holds for all $\x$ with $\|\Psi^*\f\|_0 \leq S$, where $\Psi^*$ is the conjugate transpose of $\Psi$.
\label{def:RIP_PSI}
\end{definition}


It is known that if measurement matrix $\A$ is a dense random matrix with entries chosen from independent and identically distributed (i.i.d.) symmetric Bernoulli distribution, then for any fixed sparsity basis $\Psi$, measurement matrix $\A$ satisfies RIP with high probability provided that $M \geq c S\log(N/S)$ where $M$ is the number of measurements, $N$ is the signal length, $S$ is the sparsity level of the signal and $c$ is some constant \cite{baraniuk2008simple}, \cite{candes2008Intro}.

The measurement matrix $\A$ of the proposed HFV acquisition system, however, is not a fully dense random matrix but a random block diagonal matrix.
Application of such structured measurement matrices is not limited only to our case. As mentioned in \cite{rozell2010con.rep} there are different signal acquisition settings that require such matrices either because of architectural constraints or in order to reduce computational cost of the system. For example in a network of sensors, communication constraints may force the measurements taken by each sensor dependent only on the sensor's own incident signal rather than signals from all of the sensors \cite{baron06}.  For video acquisition applying a dense random measurement matrix to the 3D data volume of an entire video incurs both high time and space complexities. A more practical approach is to take random measurements of pixels either in a temporal neighborhood (our case) or a spatial neighborhood (e.g., a fixed frame as in \cite{park2009multiscale}).
In these cases, the random measurement matrix is block diagonal rather than being a dense random matrix. Eftekhari {\it et al.} noticed the importance of such structured random measurement matrices and derived RIP of this type of matrices in \cite{eftekhari2012ris}. Based on their result, we can draw the following conclusion for our random block diagonal matrix with repeated blocks which is defined in (\ref{eq:matrix}).

\begin{theorem}
Random measurement matrix $\A$ defined in (\ref{eq:matrix}) satisfies RIP of order $S$ with isometry constant $0<\delta_S<1$ with probability $>1-2\exp(-c_1\log^2S\log^2N)$ provided that $\Psi$ is Fourier basis and $M \geq c_2 \cdot N_xN_yS\log^2S \log^2N$ where $M = K N_x N_y$, $N = T N_x N_y$, $S$ is the sparsity level of the signal; $c_1$ and $c_2$ are some constants depending only on $\delta_S$.
\label{th:rip_rbd}
\end{theorem}


Note that the above result is for the worse case scenario of random diagonal block measurement matrix, when the required number of measurements to satisfy RIP is $N_xN_yS$ times greater than conventional compressive sensing.  In the application of HFV acquisition, this worst case corresponds to the extremely rare and uninteresting HFV signal: each frame is a constant image such that all measurements taken by a camera at different pixel locations are equal.  In practice, as shown by our simulation results presented later, the HFV signal can be recovered from far fewer measurements than the bound of  Theorem~\ref{th:rip_rbd}.  Of course, compared to dense random Bernoulli measurement matrices, random measurement matrix with repeated block require more measurements to fully recover the HFV signal.
This design is solely motivated by the ease of hardware implementation of electronic shutter switch.
The 3D HFV signal can be recovered with higher fidelity for a given number of cameras $K$, if the measurement matrix $\A$ can have an increased degree of randomness than in (\ref{eq:matrix}).
In the next section we propose two superior coded HFV acquisition schemes to overcome the weakness of the initial design.

\section{Pixel-wise and Column-row-wise Coded Exposures}
\label{sec:pixel-col-row-wis}

The multi-camera coded video acquisition scheme of section \ref{sec:MCE} should be called frame-wise coded exposure, because at any given time instance $t$ and for a given camera $k$, either every or none of the $N_x N_y$ pixels of the target HFV frame $\f_t$ is exposed to camera $k$, regulated by the random exposure sequence $\bb_k$.
In this section we propose a coded HFV acquisition scheme, called pixel-wise coded exposure, to improve the performance of the frame-wise coded exposure by increasing the degree of randomness in the measurement matrix.

In pixel-wise coded exposure, camera $k$ adopts pseudo random exposure sequences that differ from pixel to pixel.  The random exposure sequence of length $T$ for pixel location $(u,v)$ is denoted by $\bb_k^{u,v}$.   Now camera $k$ modulates the time signal $\f_{u,v}$ with the random binary sequence $\bb_k^{u,v}$, and generates a random measurement of $\f_{u,v}$
\begin{equation}
y_k(u,v) = \langle \bb_k^{u,v}, \f_{u,v} \rangle + n_k .
\label{eq:pixellevel}
\end{equation}
In contrast, the frame-wise coded exposure of (\ref{eq:framelevel}) imposes the same random exposure pattern $\bb_k$ on all pixels $(u,v)$.  In the new scheme the measurement vector $\y_{u,v}$ of $\f_{u,v}$ becomes
\begin{equation}
\y_{u,v} = \B^{u,v} \f_{u,v} + \n
\end{equation}
where $\B^{u,v}$ is the $K \times T$ binary measurement matrix made of $K$ binary pseudo random sequences,
row $k$ being $\bb_k^{u,v}$, $1 \leq k \leq K$.  Therefore, the measurement matrix $\A_p$ for pixel-wise coded exposure, where $\y = \A_p \f + \n$, is
\begin{equation}
\mathbf{A}_p = \left[\begin{array}{cccc}
\mathbf{B}^{1,1} & \mathbf{0}_{K\times T} & \ldots & \mathbf{0}_{K\times T}  \\
\mathbf{0}_{K\times T} & \mathbf{B}^{1,2} & \ldots & \mathbf{0}_{K\times T}  \\
\vdots            & \vdots            &        & \vdots  \\
\mathbf{0}_{K\times T} & \mathbf{0}_{K\times T} & \ldots & \mathbf{B}^{N_x,N_y}
\end{array}\right]
\label{eq:matrix1}
\end{equation}

The measurement matrix $\A_p$ of pixel-wise coded exposure is a random block diagonal matrix with distinct and independent blocks with the entries of each block being populated from i.i.d. symmetric Bernoulli distribution. The RIP of this type of random block diagonal matrices is studied by Yap, \emph{et al.} in \cite{eftekhari2012ris}. Based on their result, we can easily derive the following conclusion for pixel-wise coded exposure.

\begin{theorem}
Let $\A_p$ be the random measurement matrix defined in (\ref{eq:matrix1}) for pixel-wise coded exposure. Matrix $\A_p$ satisfies RIP of order $S$ with isometry constant $0<\delta_S<1$ with probability $>1-2\exp(-c_1\log^2S\log^2N)$ provided that $\Psi$ is Fourier basis and $M \geq c_2\cdot S\log^2S \log^2N$ where
$c_1$ and $c_2$ are some constants depending only on $\delta_S$.
\label{th:rip_DBD}
\end{theorem}
Since as mentioned in previous section, HFV signals are sparse in frequency domain we can conclude that measurement matrix of pixel-wise coded exposure satisfies the RIP with approximately the same number of rows (within log factor) required in a dense random measurement matrix.

From the frame-wise to pixel-wise coded exposure the granularity of pixel control jumps from $O(1)$ to $O(N_x \cdot N_y)$.  This makes the complexity and cost of the latter much higher than the former.
To reduce the cost and complexity, we introduce another scheme called column-row-wise coded exposure.
In this approach we signal the $N_x$ columns and $N_y$ rows of the $N_x \times N_y$ sensor array of camera $k$ at time instance $t$ with two random binary sequences: $\rr_{k,t}$ of length $N_y$ and $\cc_{k,t}$ of length $N_x$.  The row and column binary control signals are used to produce the binary random coded exposure sequence for camera $k$ at pixel $(u,v)$ as the following
\begin{equation}
\bb_k^{u,v} = [b_k^{u,v}(1), b_k^{u,v}(2), \ldots, b_k^{u,v}(T)]'
\label{eq_colrow_bin}
\end{equation}
where $b_k^{u,v}(t) = r_{k,t}(u) \oplus c_{k,t}(v)$ and $\oplus$ is the exclusive OR operator.
In this way, only $N_x+N_y$ control signals suffice to realize pixel-wise coded exposure for each component camera of the HFV acquisition system, drastically reducing the system complexity and cost.

The measurement matrix of column-row-wise coded exposure, $\A_{cr}$, is also block diagonal with distinct blocks but diagonal blocks are not independent of each other.
However, we can have RIP for $\A_{cr}$ as stated in Theorem 3. The proof is quite involved and deferred to the appendix in order not to unduely interrupt the ongoing discussion.
\begin{theorem}
Let $\A_{cr}$ be the random measurement matrix of column-row-wise coded exposure generated from random sequence in (\ref{eq_colrow_bin}).
Matrix $\A_{cr}$ satisfies RIP of order $S$ with isometry constant $0<\delta_S<1$ with probability $>1-2\exp(-c_1\log^2S\log^2N)$ provided that $\Psi$ is Fourier basis and $M \geq c_2\cdot S\log^2S \log^2N$ where
$c_1$ and $c_2$ are some constants depending only on $\delta_S$.
\label{th:rip_DBD:Acr}
\end{theorem}

Theorems 2 and 3 conclude that the pixel-wise and column-row-wise coded exposures asymptotically require  the same number of random measurements to recover the HFV signal; they both perform almost as well as the dense random measurement matrix within a logarithm factor.  In practice,
the column-row-wise coded exposure should be the method of choice for it is simpler and more cost effective to implement, without sacrificing the performance compared with the pixel-wise coded exposure.


\section{Sparsity-based HFV Recovery}
\label{sec:HFV_Recovery}

The recovery model defined in (\ref{eq:min:sparse:x}) is usually called \emph{sparse synthetic model} which now has solid theoretical foundations and is a stable field \cite{li2012note}.
Alongside this approach there is \emph{sparse analysis model} which uses a possibly redundant analysis operator $\Theta \in \mathds{R}^{P\times N}$ ($P\geq N$) to exploit sparsity of signal $\f$, i.e., signal $\f$ belongs to analysis model if $\|\Theta\f\|_0$ is small enough \cite{Nam2013cosparse}.

In this section we develop a HFV recovery method based on a sparse analysis model which
employs
strong temporal and spatial correlations of the HFV signal.
If we assume that the object(s) in the video scene has flat surfaces and is illuminated by parallel light source like the sun, then the 2D intensity function $\f_t(u,v)$ of frame $t$ can be approximated by a piecewise constant function based on the Lambatian illumination model.  For the same reason we can use a piecewise linear model of $\f_t(u,v)$ if the light source is close to the object or/and the objects in the scene have surfaces of small curvature.  Assuming that each target frame $\f_t(u,v)$ is a 2D piecewise linear function, the Laplacian $\nabla^2_{u,v} \f_t$ of $\f_t(u,v)$ is zero or near zero at most pixel positions of the $uv$ plane and takes on large magnitudes only at object boundaries and texture areas.  In other words, $\nabla^2_{u,v} \f_t$ offers a sparse representation of $\f_t$ that results from intra-frame spatial correlations.

The other source of sparsity is rooted in temporal correlations of HFV. Precisely because of high frame rate, the object motion between two adjacent frames is very small in magnitude so that most pixels will remain in the same object from frame $\f_t$ to $\f_{t+1}$.  Also, general affine motion can be
satisfactorily approximated by translational motion $(du, dv)$ if it is small enough.  As long as the small motion $(du, dv)$ does not move a pixel outside of an object whose intensity function is linear, i.e.,
$f_t(u,v) = au + bv + c$, we have
\begin{eqnarray}
\nabla_t f_t(u,v) &=& f_{t+1}(u,v) - f_t(u,v)  \nonumber \\
&=& f_t(u+du, v+dv) - f_t(u,v)  \nonumber \\
&=& a du + b dv .
\label{eq:const}
\end{eqnarray}
This means that the first-order difference $\nabla_t f_t(u,v)$ in time remains constant in the intersection region of an object segment across two adjacent frames.  By considering $\nabla_t \f_t$ as a 2D function in the $uv$ plane, it follows from  (\ref{eq:const}) that $\nabla_t \f$ is piecewise constant.  Therefore, the total variation of the 2D function $\nabla_t \f$, namely
$\nabla_{u,v} (\nabla_t \f)$, is another sparse representation of $\f$.

Using the two sparsity models described above we can now define the redundant analysis operator, $\Theta$ of size $(2TN_xN_y)\times(TN_xN_y)$, as
\[
\Theta  =
  \begin{bmatrix}
       \Theta_1\\
       \Theta_2
  \end{bmatrix}
\]
where $\Theta_1$ of size $(TN_xN_y)\times(TN_xN_y)$ and $\Theta_2$ of size $(TN_xN_y)\times(TN_xN_y)$ are matrix representations of the Laplacian operator ($\nabla^2_{u,v}$) and $\nabla_{u,v}(\nabla_t)$ operator respectively.
Since, as explained above, $\Theta \f$ is sparse, we can consider $\Theta$ to be the redundant analysis operator for HFV signal $\f$.
Therefore, we can formulate the recovery of $\f$ from (\ref{eq:underdet}) in the context of this sparse analysis model as
\begin{equation}
\min_{\f} \|\Theta \f\|_1 \;\;\;\; s.t. \;\;\;\; \|\A \f - \y\|_2 \leq \sigma
\label{eq_3dcs}
\end{equation}
where $\sigma$ is the variance of the measurement error.

In general, sparse synthesis model defined in (\ref{eq:min:sparse:x}) and above sparse analysis model are different. In a special case where $\Theta$ is orthonormal, the two models are the same with $\Theta = \Psi^{-1}$ \cite{Nam2013cosparse}.  Although
a large number of applications for (\ref{eq_3dcs}) are found, theoretical study of sparse analysis model is not as thorough and well established as sparse synthesis model in the compressive sensing literature.
Recently, Li in \cite{li2012note} addressed this gap by introducing the following generalized RIP.
\begin{definition}
Measurement matrix $\A$ satisfies generalized RIP of order $S$ with isometry constant $0<\delta_S<1$ if
\[
(1-\delta_S)\|\Theta\f\|_2^2 \leq \|\A\f\|_2^2 \leq (1+\delta_S) \|\Theta\f\|_2^2
\]
holds for all $\f$ which are $S$-sparse after transformation of $\Theta$, i.e., $\|\Theta\f\|_0 \leq S$.
\end{definition}
Li shows that if measurement matrix $\A$ satisfies generalized RIP with $\delta_{2S} < \sqrt{2}-1$, it is guaranteed that the sparse analysis model defined in (\ref{eq_3dcs}) can accurately recover signal $\f$ which is sparse in arbitrary overcomplete and coherent operator $\Theta$. It is also shown that nearly all random matrices that satisfy RIP will also satisfy generalized RIP.
However to the best of our knowledge, no theoretical results have been published yet that prove generalized RIP for the random measurement matrices that are employed here.

Although there is so far no theoretical performance result for sparse analysis model when applied to the recovery of HFV signals with random block diagonal matrices, the experimental results (section~\ref{sec:sim_res}) show that, given the number of cameras, the quality of recovered HFV signals using (\ref{eq_3dcs}) is better than the quality of recovered HFV signals using the sparse synthesis model defined in (\ref{eq:min:sparse:x}).


We would like to end this section with a remark on the possibility of more sophisticated
HFV recovery algorithms.  The above introduced method is effective under the assumption of
linear spatial and temporal correlations.  But other forms of sparsity of the HFV signal, say, sparsity in a transform domain (e.g., spaces of DWT, PCA, etc.), can be readily exploited very much the same ways as in the large body of literature on image restoration and compressive sensing.  The focus of this work is, however, on coded HFV acquisition.

\section{System Advantages}
\label{sec:system_attributes}

%

The proposed sparsity-based HFV recovery algorithm from random measurements has an important side benefit:
the relative simplicity of cameras assembly and calibration compared with other multicamera systems such as the one in \cite{WilburnJoshiVaish04}.  After the $K$ cameras are compactly mounted, the relative displacements among these cameras can be measured by imaging a calibration pattern.  But there is no need for precise spatial registration of all the $K$ cameras.  On the contrary, random perturbations of the pixel grids of different cameras are beneficial to the HFV recovery because they add a degree of randomness to the measurements of coded exposure in spatial domain, as explained below.

\begin{figure}[!t]
\centering
\begin{minipage}{\linewidth}
\centering
\subfloat[]{\includegraphics[width=.5\linewidth]{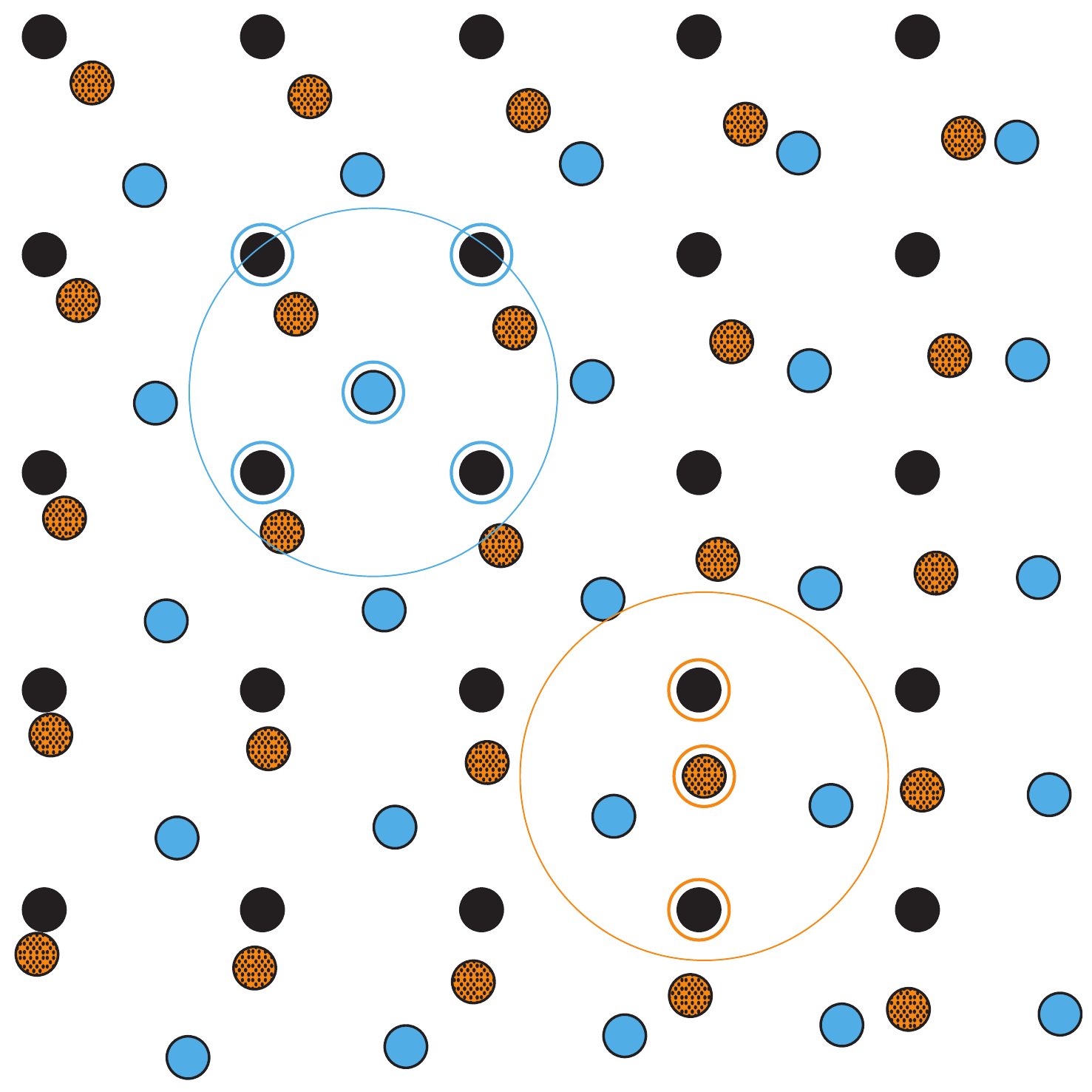}}%
\hfil
\subfloat[]{\includegraphics[width=.5\linewidth]{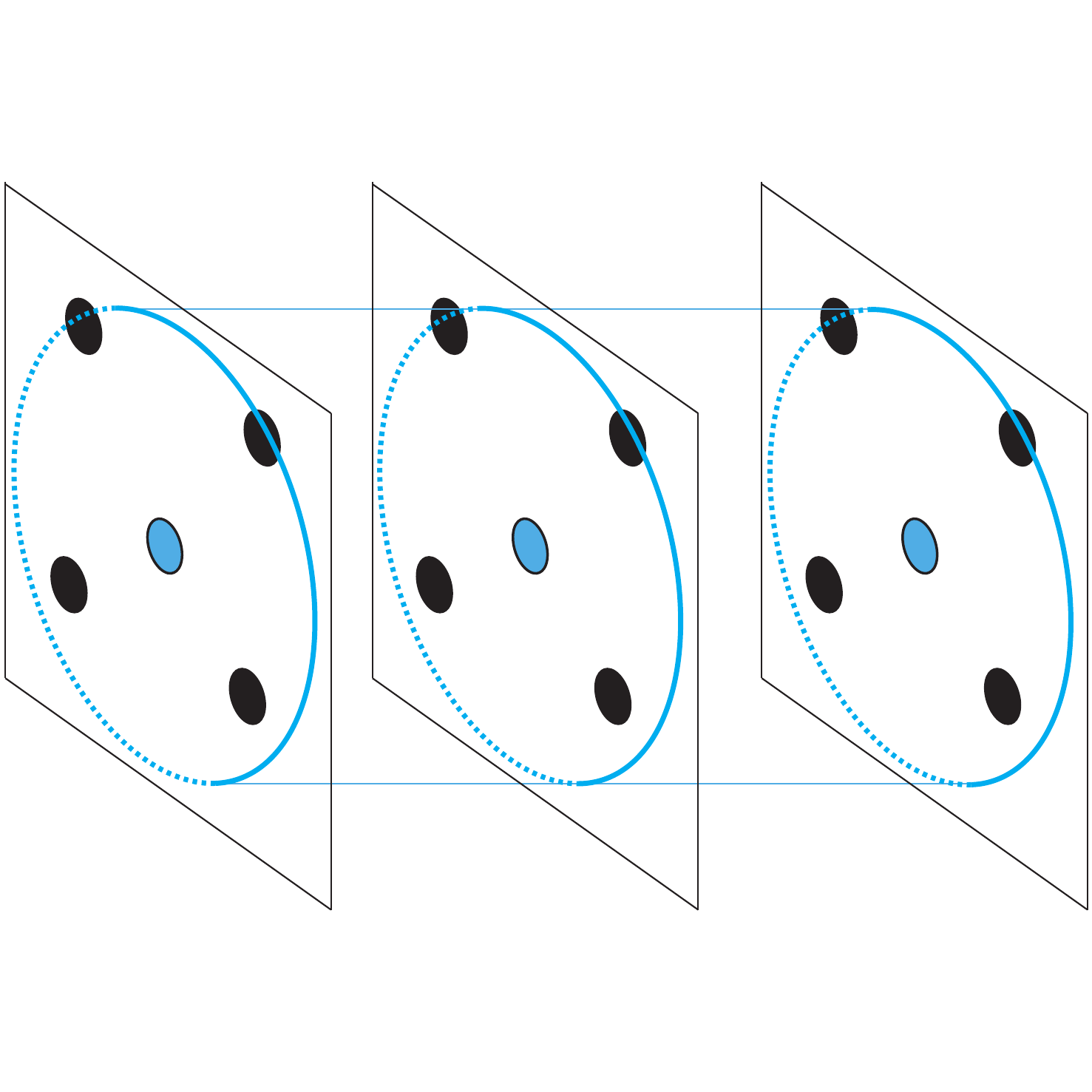}}%
\end{minipage}
\caption{(a) Relative positioning of the cameras in the image plane. Black circles are the pixels of HFV signal we want to reconstruct. Blue and orange circles are the pixels captured by different cameras. (b) Cylindrical view of an specific pixel of a camera in time.}
\label{fig:random_layout}
\end{figure}

As illustrated in Fig.~\ref{fig:random_layout}, once the relative positioning of the cameras in the image plane is determined and given the point spread function (PSF) of the cameras, each of the $K N_x N_y$ random measurements of the HFV signal $\f$ can be viewed, via coded acquisition, as a random projection of the pixels in a cylindrical spatial-temporal neighborhood.

The pixel value recorded by camera $k$ at location $(x,y)$ and time $t$ is
\begin{equation}
g_t^{k}(x,y) = \sum_{(u,v)\in W} f_t(u,v)h(x-u,y-v)
\label{eq:conv}
\end{equation}
where $h(\cdot,\cdot)$ is the PSF of the camera with convolution window $W$.
By forming a super vector $\g$ out of all the $K T N_x N_y$ pixels captured by the $K$ cameras the same way as in $\f$, we can write (\ref{eq:conv}) in matrix form as
\begin{equation}
\g = \Ha \f
\label{eq:conv:mtx}
\end{equation}
where $\Ha$ is the $K T N_x N_y \times T N_x N_y$ convolution matrix that is determined by the relative positioning of the cameras and the camera PSF.  Each row of $\Ha$ consists of the weights $h(x-u,y-v)$ in (\ref{eq:conv}) for distinctive $k$ and $(x,y)$.

Given pixel location $(x,y)$, camera $k$ modulates the time signal $\g^k_{x,y} = (g_1^{k}(x,y), g_2^{k}(x,y), \cdots, g_T^{k}(x,y))'$ by the binary coded exposure sequence $\bb_k^{x,y}$, and generates a random measurement of $\f_{u,v}$'s, $(u,v) \in W$,
\begin{equation}
y_k(x,y) = \langle \bb_k^{x,y}, \g^k_{x,y} \rangle + n_k .
\label{eq:measurePSF}
\end{equation}
Using super vector $\g$, we can also represent (\ref{eq:measurePSF}) in the matrix form as
\begin{equation}
\y = \B \g + \n .
\label{eq:measurePSF:mtx}
\end{equation}
where $\B$ is the $K N_x N_y \times K T N_x N_y$ binary matrix made of $K N_x N_y$ binary pseudo random sequences as follows
\begin{equation*}
\B = \left[\begin{array}{cccccccccc}
\bb_1^{1,1} & \ldots & \mathbf{0}_{1\times T} & \ldots & \mathbf{0}_{1\times T} & \ldots & \mathbf{0}_{1\times T} \\
\vdots & & \vdots & & \vdots & & \vdots \\
\mathbf{0}_{1\times T} & \ldots & \bb_K^{1,1} & \ldots & \mathbf{0}_{1\times T} & \ldots & \mathbf{0}_{1\times T} \\
\vdots & & \vdots & & \vdots & & \vdots \\
\mathbf{0}_{1\times T} & \ldots & \mathbf{0}_{1\times T} & \ldots & \bb_k^{i,j} & \ldots & \mathbf{0}_{1\times T} \\
\vdots & & \vdots & & \vdots & & \vdots \\
\mathbf{0}_{1\times T} & \ldots & \mathbf{0}_{1\times T} & \ldots & \mathbf{0}_{1\times T} & \ldots & \bb_K^{N_x,N_y}
\end{array}\right]
\label{eq:matrix_BIN:PSF}
\end{equation*}
We can combine the operations (\ref{eq:conv:mtx}) and (\ref{eq:measurePSF:mtx}) and express the $K N_x N_y$ random measurements in the following matrix form
\begin{eqnarray*}
\y = \B \g +\n = \B \Ha \f +\n= \A \f+\n
\end{eqnarray*}
where $\A$ is the $K N_x N_y \times T N_x N_y$ random measurement matrix, which can be determined once the relative positioning of the $K$ cameras is measured via the calibration of camera assembly.




%

The ability of the proposed multicamera system to acquire very high speed video with conventional cameras is at the expense of very high complexity of the HFV reconstruction algorithm.  In this system aspect, the new HFV acquisition technique seems, on surface, similar to compressive sensing.  But the former can be made computationally far more practical than the latter, thanks to a high degree of parallelism in the solution of (\ref{eq_3dcs}).  Recalling from the previous discussions and Fig.~\ref{fig:random_layout}, a random measurement of the 3D signal $\f$ made by coded exposure is a linear combination of the pixels in a cylindrical spatial-temporal neighborhood.  Therefore, unlike in compressive sensing for which the signal $\f$ has to be recovered as a whole via $\ell_1$ minimization, our inverse problem (\ref{eq_3dcs}) can be broken into subproblems;  a large number of $W \times H \times T$ 3D sample blocks, $W < N_x$, $H < N_y$, can be processed in parallel, if high speed recovery of HFV is required.
Particularly worth noting is that partitioning of the problem (\ref{eq_3dcs}) into $J$ parts can potentially  speed up the HFV reconstruction by a factor of $O(J^3)$, far more than $J$ folds.  This is because the time complexity of linear programming for $\ell_1$ minimization is cubical in the problem size.

Separately solving (\ref{eq_3dcs}) for $W \times H \times T$ blocks does not compromise the quality of recovered HFV as long as the 2D image signal is sparse in the $W \times H$ cross section of the block.  On the contrary, this strategy can improve the quality of recovered HFV if overlapped domain blocks are used when solving (\ref{eq_3dcs}).  If a pixel is covered by $m$ such domain blocks, then the solutions of the multiple instances of linear programming, one per domain block, yield $m$ estimates of the pixel.  These $m$ estimates can be fused to generate a more robust final estimate of the pixel.


\section{Simulation Results}
\label{sec:sim_res}

We begin this section by first showing that HFV signals are actually sparse in frequency domain. Next, we compare the performance of proposed schemes (frame-wise, pixel-wise and column-row-wise) in the context of sparse synthesis model defined in (\ref{eq:min:sparse:x}). After that we compare the performance of sparse synthesis model against sparse analysis model for pixel-wise coded exposure scheme. In the reminder of this section we report simulation results of the proposed multicamera HFV acquisition techniques based on sparse analysis, and compare the performances of the proposed coded HFV acquisition schemes.

\begin{figure}[!t]
\centering
\includegraphics[width=3.0in]{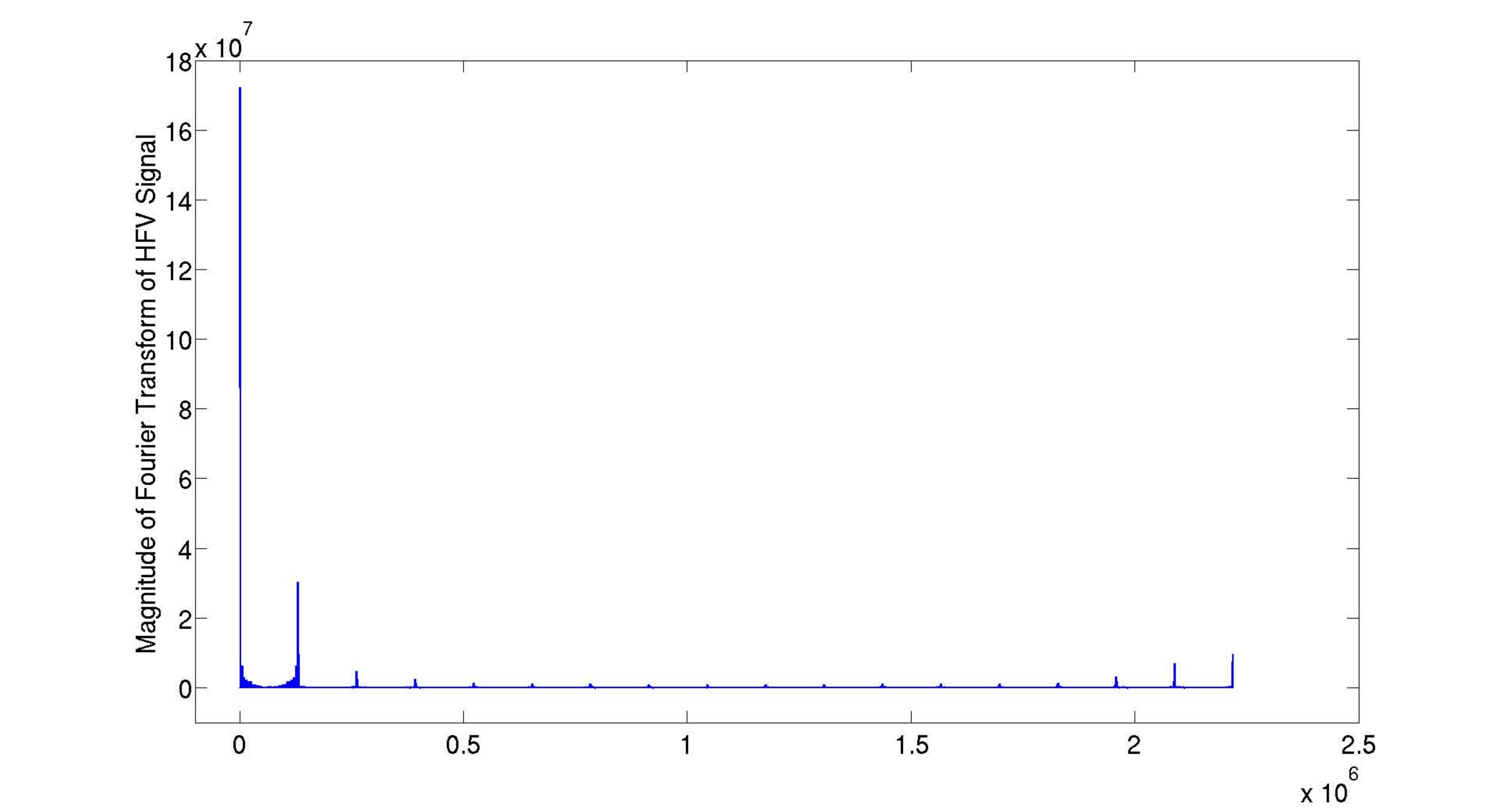}
\caption{Magnitude of DFT coefficients of the HFV sequence ``Airbag'', clearly showing the sparsity of the HFV signal in Fourier domain.}
\label{fig:hfv_fourier_transform}
\end{figure}

To show that HFV signals are indeed sparse in frequency domain, we applied 3-dimensional Fourier transform on different HFV sequences. As expected, the DFT coefficients of all sample HFV signals have only few large values and the remainders are zero or near zero. To show the level of sparsity of HFV signals in the frequency domain, we plot in Fig.~\ref{fig:hfv_fourier_transform} the magnitude of the DFT coefficients for the HFV sequence `Airbag' as an example.  Since HFV signals have sparse representation in frequency domain, we can use the sparse synthesis model defined in (\ref{eq:min:sparse:x}) with $\Psi$ being matrix form of 3-dimensional Fourier transform to recover HFV signals from (\ref{eq:underdet}).

\begin{figure*}[!t]
\centering
\begin{minipage}{\linewidth}
\centering
\subfloat[]{\includegraphics[trim=207 37 187 54, clip,width=1.5in]{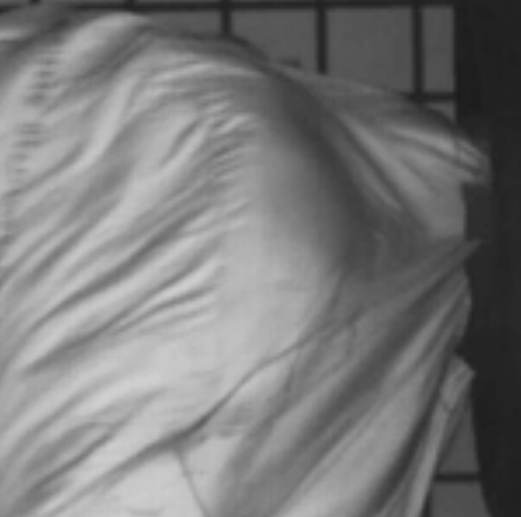}}%
\hfil
\subfloat[]{\includegraphics[trim=207 37 187 54, clip,width=1.5in]{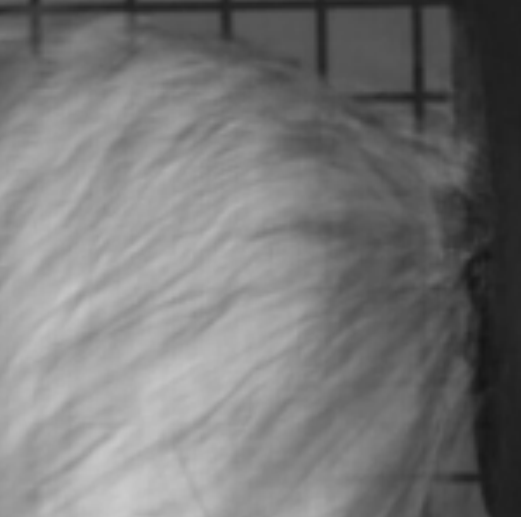}}%
\hfil
\subfloat[]{\includegraphics[trim=207 37 187 54, clip,width=1.5in]{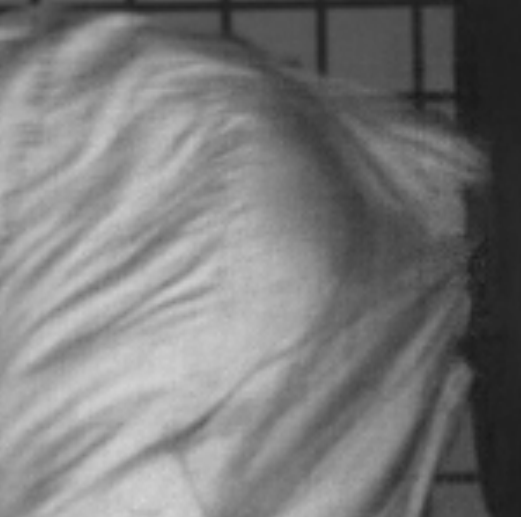}}%
\hfil
\subfloat[]{\includegraphics[trim=207 37 187 54, clip,width=1.5in]{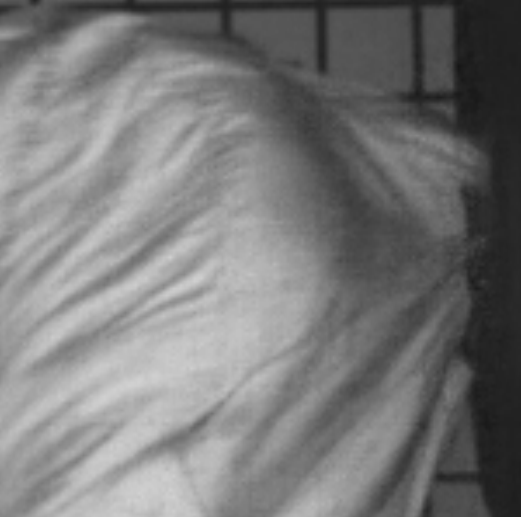}}%
\end{minipage}
\caption{Performance comparison of the 8-cameras frame-wise, pixel-wise and column-row-wise coded exposures with sparse synthesis-based recovery in Fourier domain. (a) An original frame. (b) Recovered frame by frame-wise coded exposure. Average PSNR of all recovered frames is 25.60dB. (c) Recovered frame by the pixel-wise coded exposure. Average PSNR of all recovered frames is 41.26dB. (d) Recovered frame by the column-row-wise coded exposure. Average PSNR of all recovered frames is 41.26dB.}
\label{fig:cmp_fourier:pixelwise}
\end{figure*}

To compare the performance of proposed schemes (frame-wise, pixel-wise and column-row-wise coded exposures) in the context of sparse synthesis model defined in (\ref{eq:min:sparse:x}), we present in Fig.~\ref{fig:cmp_fourier:pixelwise} the snapshots of recovered ``Airbag'' HFV signal as well as average PSNRs of all recovered frames. As previously stated in section~\ref{sec:MCE}, when using the random measurement matrix of frame-wise coded exposure, we lose the effectiveness of measurements compared to dense random measurement matrix by a factor of $>N_xN_y$. This loss of effectiveness is apparent in Fig.\ref{fig:cmp_fourier:pixelwise}: with the same number of cameras, the quality of recovered HFV signal for frame-wise coded exposure is not as good as other schemes. From this figure we can also see that the quality of recovered HFV signals for pixel-wise and column-row-wise coded exposure schemes are almost the same.

\begin{figure}[!t]
\centerline{\subfloat[]{\includegraphics[trim=207 37 187 54, clip,width=1.5in]{airbag1000fps_240x544_jFFT_bin_1003T59_gl_pb_cams8_sh1_s0_f14}%
\label{fig_first_case}}
\hfil
\subfloat[]{\includegraphics[trim=207 37 187 54, clip,width=1.5in]{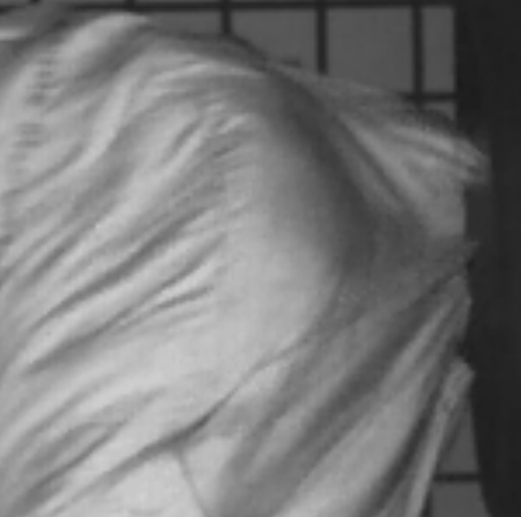}%
\label{fig_second_case}}}
\caption{Comparison of the performance of sparse synthesis and analysis models with pixel-wise coded exposure for HFV sequence ``Airbag'' with 8 cameras. (a) Snapshot of recovered HFV sequence using sparse synthesis model. Average PSNR of recovered frames is 41.26dB. (b) Snapshot of recovered HFV sequence using sparse analysis model. Average PSNR of recovered frames is 44.49dB.}
\label{fig:cmp:synthesis:analysis}
\end{figure}

In Fig.~\ref{fig:cmp:synthesis:analysis} we compare the performance of sparse synthesis and analysis models and present the snapshots of recovered HFV signals as well as average PSNRs of all recovered frames. Since column-row-wise coded exposure performs almost the same as pixel-wise coded exposure (see Fig.~\ref{fig:cmp_fourier:pixelwise}), in this figure we only represent the results of using pixel-wise coded exposure with the two models. As it can be seen, the analysis model performs much better compared to the synthesis model.

In the rest of this section, we report simulation results of the proposed multicamera HFV acquisition techniques based on the sparse analysis model defined in (\ref{eq_3dcs}), and compare the performances of three different modes of coded HFV acquisition: frame-wise, pixel-wise and column-row-wise coded exposures.
Since the measurements matrices of proposed techniques are random, the quality of reconstructed HFV signal may vary over different random coded acquisition sequences.  In order to make our findings statistically significant, for each of the three HFV acquisition methods and each high-speed test video we conducted 75 experiments with different random coded acquisition sequences, and calculated the average of these results. In Table~\ref{tbl:av_psnr}, average PSNR values of six reconstructed
HFV sequences: ``Airbag" (1000 fps), ``Apple Cutting" (2000 fps), ``Car Crash" (1000 fps), ``Bird" (1000 fps), ``Bulb Bursting 1'' (2000 fps) and ``Bulb Bursting 2'' (2000fps), are tabulated for different number of cameras (each has a frame rate of 60 fps) and for the three different modes of coded exposure.
As demonstrated in Table~\ref{tbl:av_psnr}, the performance of pixel-wise and column-row-wise coded exposure techniques are almost the same and both perform significantly better than frame-wise coded exposure technique.

\begin{table}[!t]
\caption{Average PSNR of recovered HFV using different number of cameras.}
\label{tbl:av_psnr}
\centering
\begin{tabular}{cc|cccc|}
\cline{3-6}
& & \multicolumn{4}{|c|}{Number of Cameras} \\ \cline{3-6}
& & 4 & 8 & 12 & 16 \\ \cline{1-6}
\multicolumn{1}{|c|}{\multirow{3}{*}{Airbag}} &
\multicolumn{1}{|c|}{Frame-wise} & 24.17 & 25.40 & 29.38 & 43.18  \\ 
\multicolumn{1}{|c|}{}                        &
\multicolumn{1}{|c|}{Pixel-wise} & 39.52 & 44.54 & 48.85 &  54.62  \\ 
\multicolumn{1}{|c|}{}                        &
\multicolumn{1}{|c|}{Column-row-wise} & 39.55 & 44.57 & 48.90 & 54.61  \\ \cline{1-6}
\multicolumn{1}{|c|}{\multirow{3}{*}{Apple Cutting}} &
\multicolumn{1}{|c|}{Frame-wise} & 24.15 & 25.56 & 26.59 & 28.28  \\ 
\multicolumn{1}{|c|}{}                        &
\multicolumn{1}{|c|}{Pixel-wise} & 39.51 & 44.02 & 46.48 & 48.64  \\ 
\multicolumn{1}{|c|}{}                        &
\multicolumn{1}{|c|}{Column-row-wise} & 39.53 & 44.04 & 46.47 & 48.63  \\ \cline{1-6}
\multicolumn{1}{|c|}{\multirow{3}{*}{Car Crash}} &
\multicolumn{1}{|c|}{Frame-wise} & 25.95 & 27.92 & 32.52 & 44.78 \\ 
\multicolumn{1}{|c|}{}                        &
\multicolumn{1}{|c|}{Pixel-wise} & 39.41 & 46.08 & 51.84 & 58.31 \\ 
\multicolumn{1}{|c|}{}                        &
\multicolumn{1}{|c|}{Column-row-wise} & 39.34 & 46.10 & 52.00 & 58.30 \\ \cline{1-6}
\multicolumn{1}{|c|}{\multirow{3}{*}{Bird}} &
\multicolumn{1}{|c|}{Frame-wise} & 31.98 & 34.21 & 39.71 & 51.82 \\ 
\multicolumn{1}{|c|}{}                        &
\multicolumn{1}{|c|}{Pixel-wise} & 40.52 & 47.22 & 53.89 & 61.92 \\ 
\multicolumn{1}{|c|}{}                        &
\multicolumn{1}{|c|}{Column-row-wise} & 40.49 & 47.34 & 53.76 & 61.77 \\ \cline{1-6}
\multicolumn{1}{|c|}{\multirow{3}{*}{Bulb Bursting 1}} &
\multicolumn{1}{|c|}{Frame-wise} & 32.23 & 34.38 & 35.98 & 38.14 \\ 
\multicolumn{1}{|c|}{}                        &
\multicolumn{1}{|c|}{Pixel-wise} & 35.62 & 39.18 & 41.72 & 43.97 \\ 
\multicolumn{1}{|c|}{}                        &
\multicolumn{1}{|c|}{Column-row-wise} & 35.71 & 39.21 & 41.75 & 43.92  \\ \cline{1-6}
\multicolumn{1}{|c|}{\multirow{3}{*}{Bulb Bursting 2}} &
\multicolumn{1}{|c|}{Frame-wise} & 36.01 & 37.52 & 38.98 & 40.43  \\ 
\multicolumn{1}{|c|}{}                        &
\multicolumn{1}{|c|}{Pixel-wise} & 38.97 & 42.56 & 45.09 & 47.73  \\ 
\multicolumn{1}{|c|}{}                        &
\multicolumn{1}{|c|}{Column-row-wise} & 38.93 & 42.57 & 45.11 & 47.65 \\ \cline{1-6}
\end{tabular}
\end{table}

\begin{figure}[t]
\centering
\begin{minipage}{\linewidth}
\centering
\subfloat[``Apple Cutting", $8$ cameras.]{\includegraphics[width=.5\linewidth]{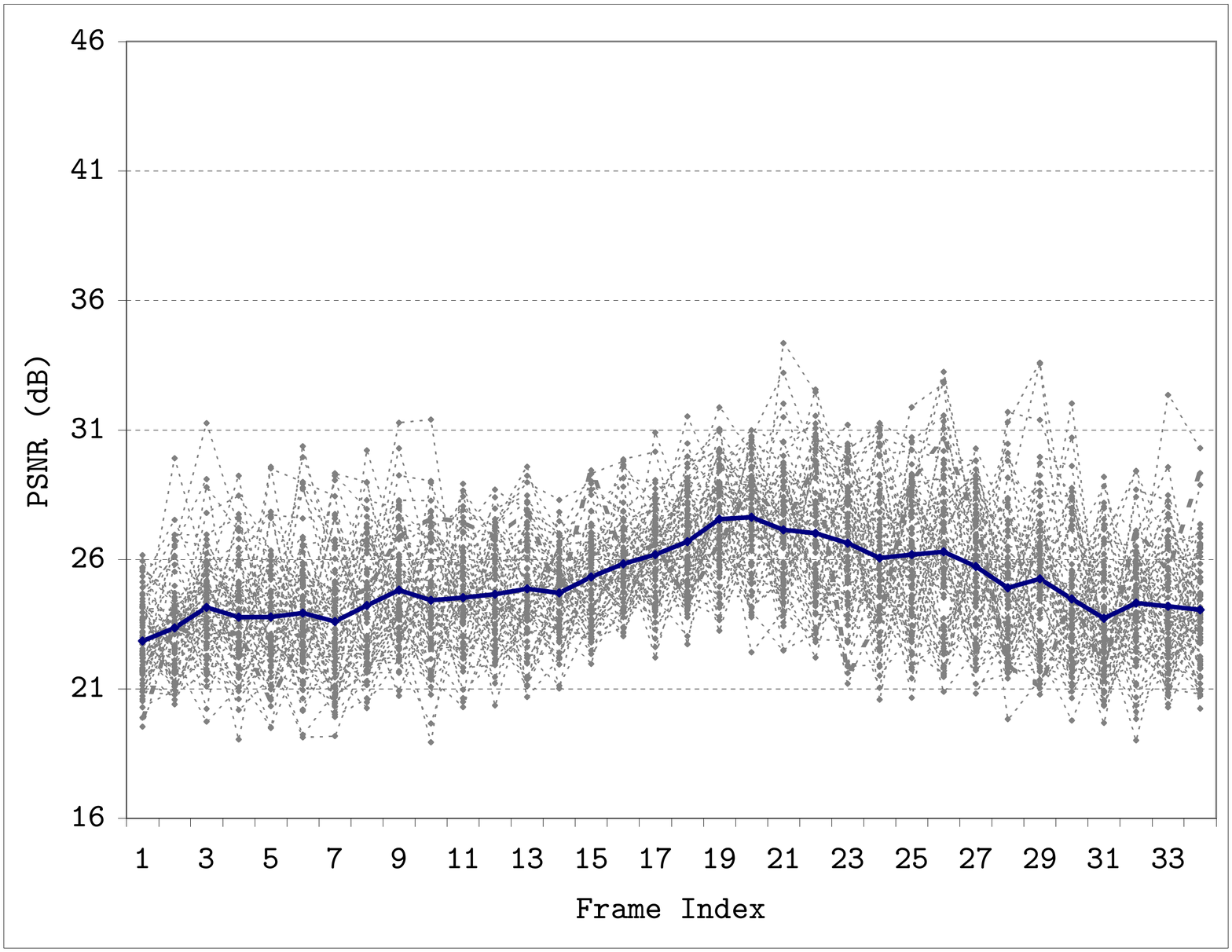}}%
\hfil
\subfloat[``Apple Cutting", $8$ cameras.]{\includegraphics[width=.5\linewidth]{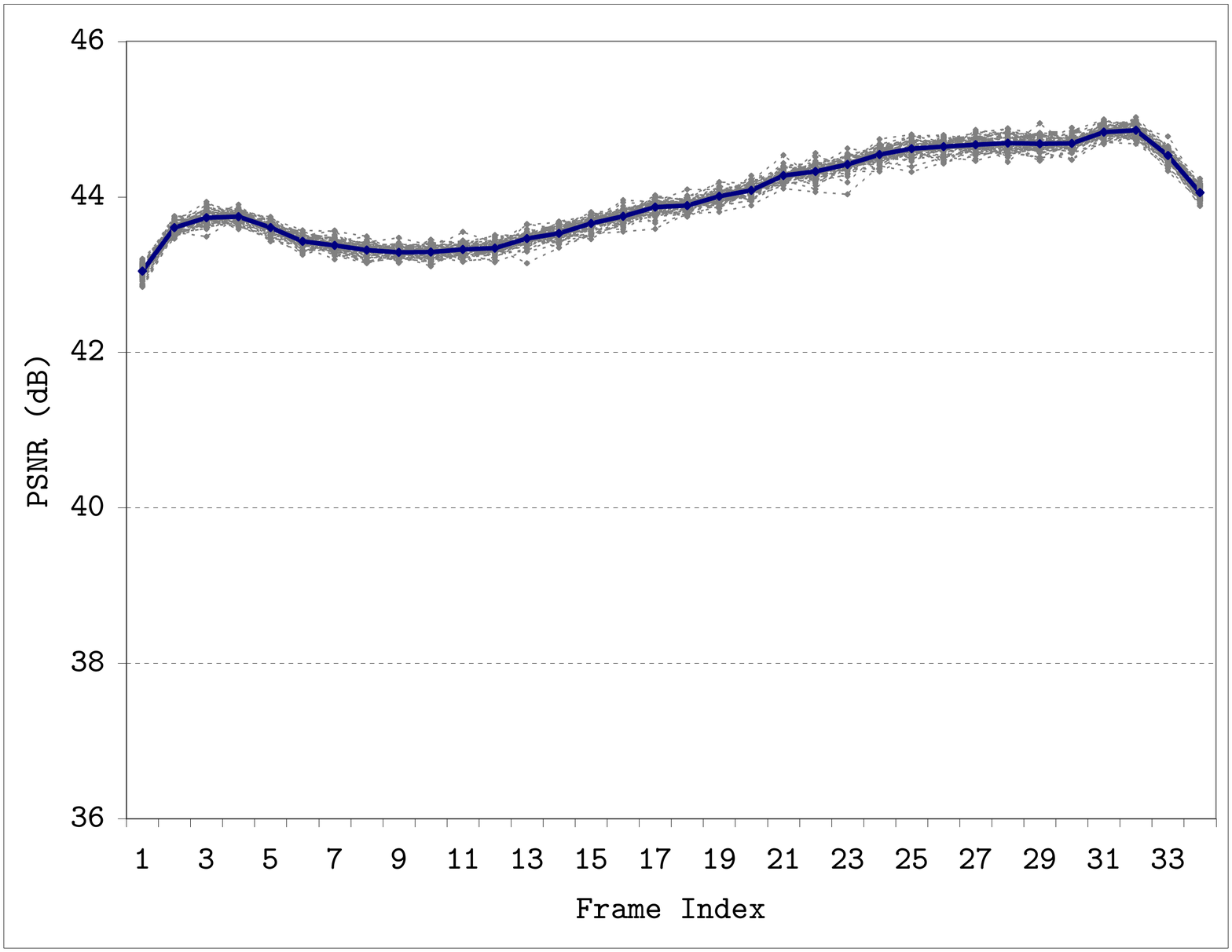}}%
\end{minipage}
\begin{minipage}{\linewidth}
\centering
\subfloat[``Bird", $12$ cameras.]{\includegraphics[width=.5\linewidth]{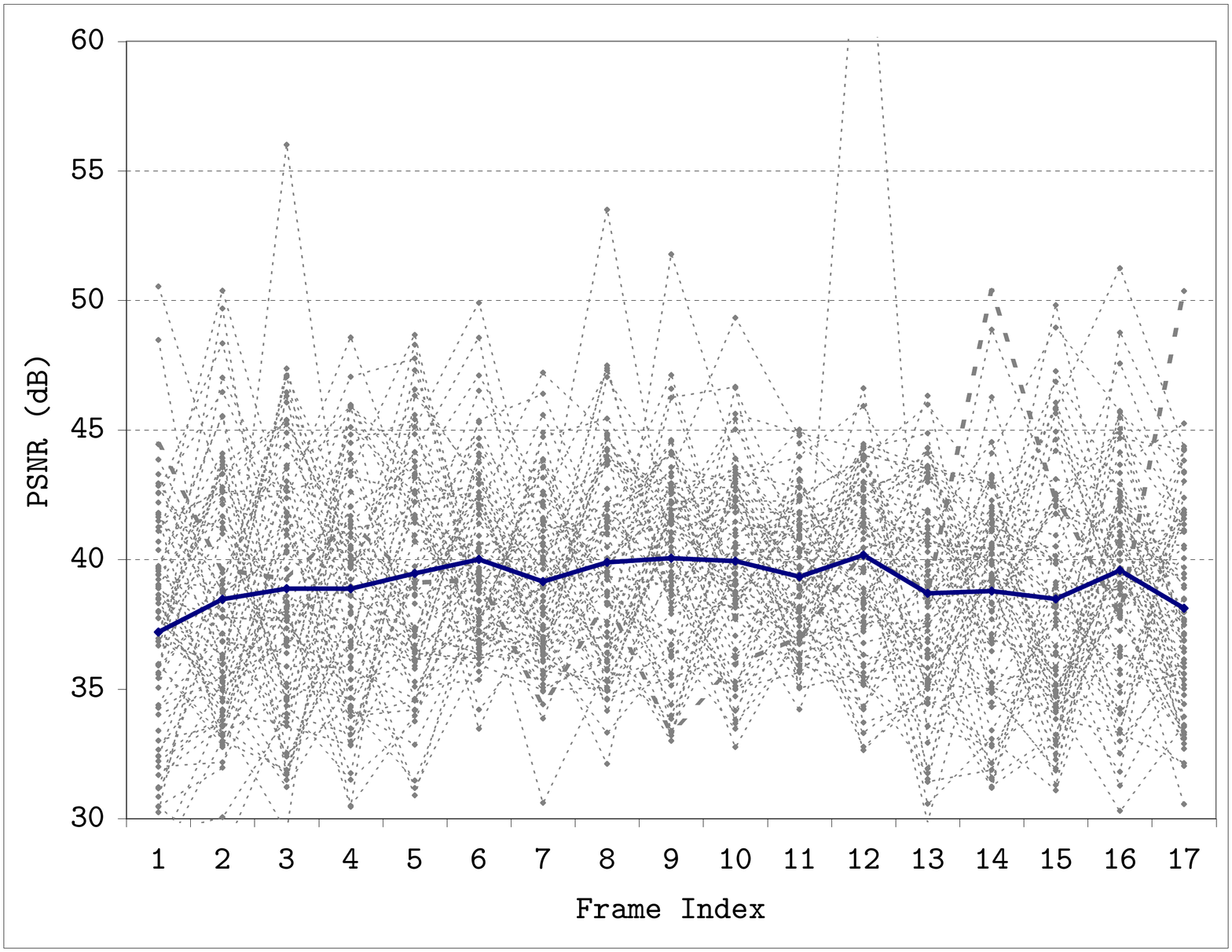}}%
\hfil
\subfloat[``Bird", $12$ cameras.]{\includegraphics[width=.5\linewidth]{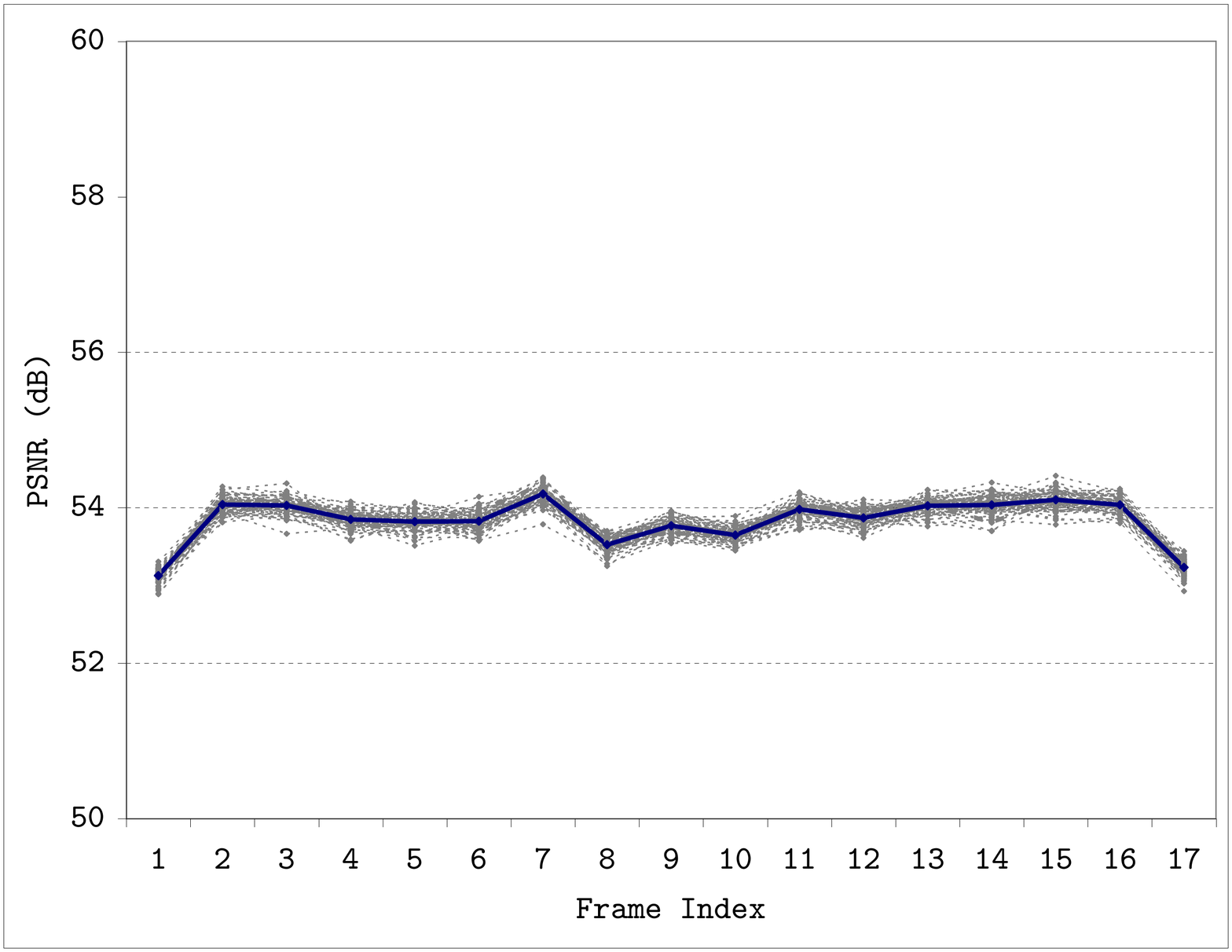}}%
\end{minipage}
\caption{Family of 75 PSNR curves (one per random coded acquisition sequence) with respect to frame index.
Left: Frame-wise coded acquisition;
Right: Column-row-wise coded acquisition (the pixel-wise coded acquisition has virtually same curves).
}
\label{fig:multi_run}
\end{figure}

Fig.~\ref{fig:multi_run} represents the PSNR curves of individual frames for 75 runs of the experiments on two high-speed test video sequences. A gray line in the plots represents the PSNR curve of a different run and the blue curve is the average of the 75 PSNR curves.  Experiments show that the column-row-wise and pixel-wise coded acquisition techniques perform virtually the same for all HFV frames.  To save the space, only the curves of column-row-wise coded exposure are plotted in Fig.~\ref{fig:multi_run} and in all other figures of this section.  As demonstrated in Fig.~\ref{fig:multi_run}, column-row-wise and pixel-wise coded acquisitions are very robust; the quality of the reconstructed HFV signals does not depend on the random coded acquisition sequence that is used. But this is not the case for the frame-wise coded acquisition; its performance oscillates wildly over different random coded acquisition sequences.

\begin{figure}[!t]
\centering
\begin{minipage}{\linewidth}
\centering
\subfloat[``Car Crash"]{\includegraphics[width=.5\linewidth]{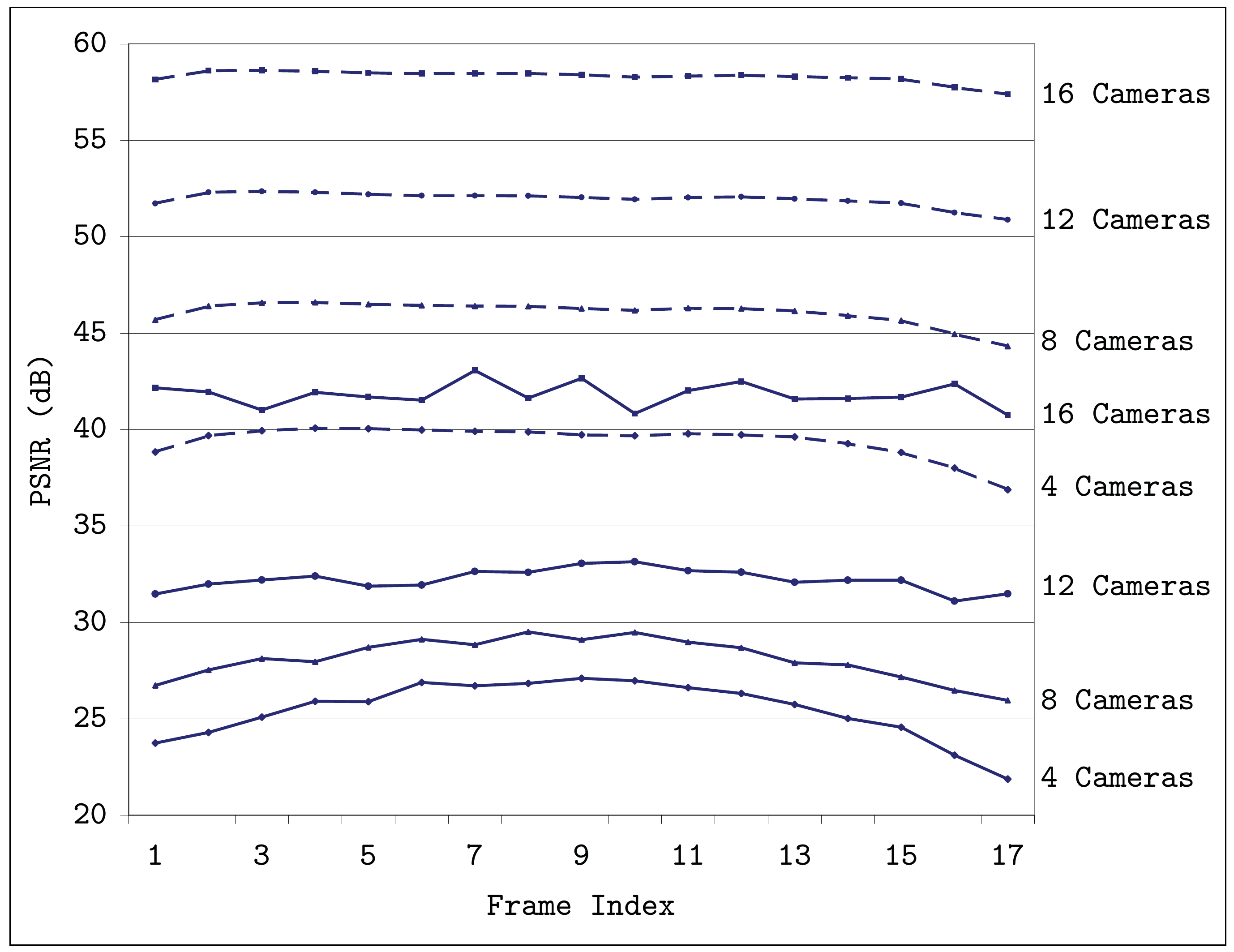}}
\hfil
\subfloat[``Apple Cutting"]{\includegraphics[width=.5\linewidth]{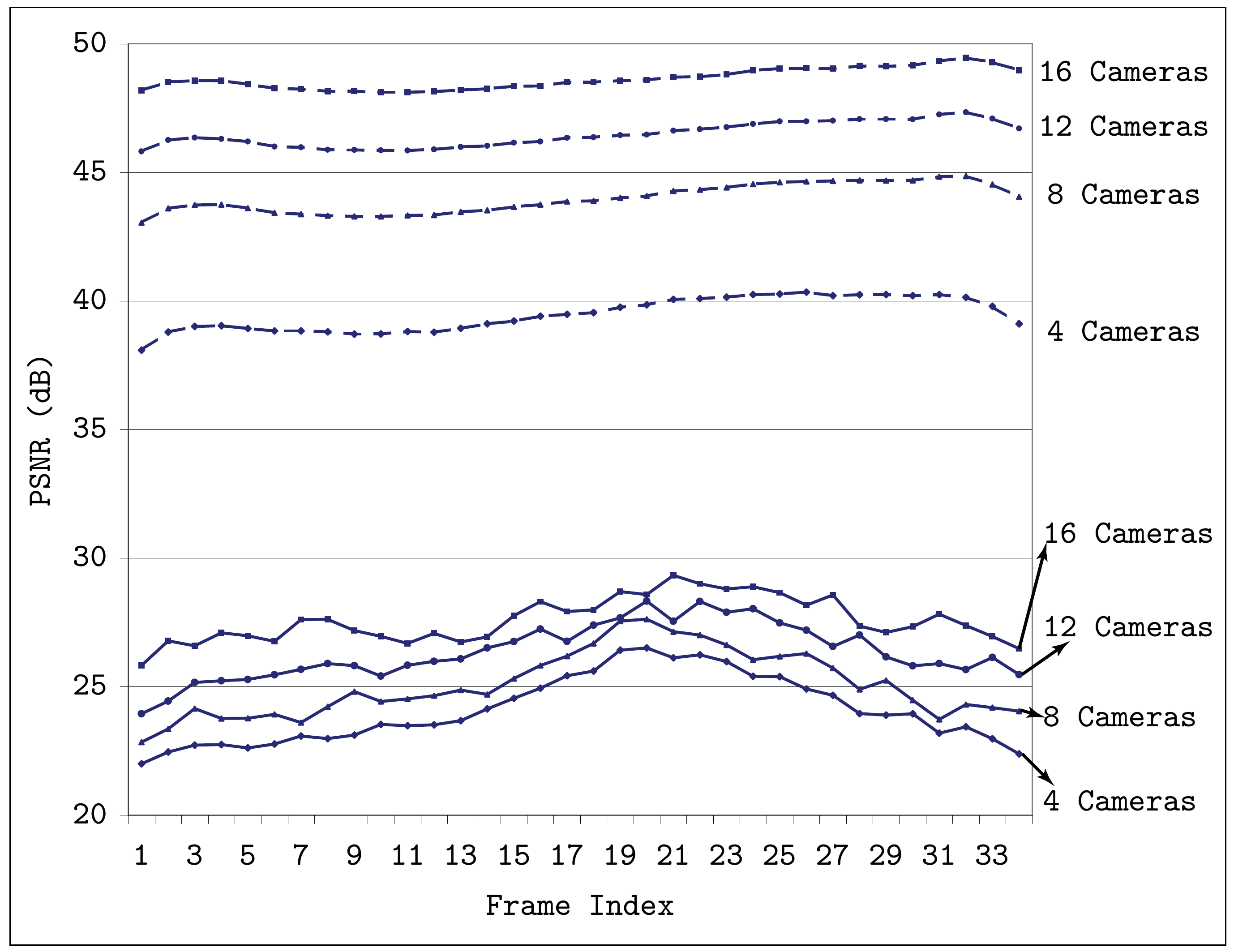}}
\end{minipage}
\caption{Performance comparison of frame-wise and column-row-wise code exposure techniques using different number of cameras (solid-lines represent frame-wise PSNR curves; dashed-lines represent column-row-wise PSNR curves).}
\label{fig:psnr:fl_cl_all}
\end{figure}

Fig.~\ref{fig:psnr:fl_cl_all} compares the performances of the frame-wise and column-row-wise coded HFV acquisitions with respect to different number of cameras.
As demonstrated, for a fixed number of cameras, the quality of target frames recovered by column-row-wise acquisition is steady (almost flat PSNR curves), whereas the corresponding PSNR curves of the frame-wise coded acquisition fluctuate by as much as 4dB between the recovered frames.  This unpredictable behavior of the frame-wise coded acquisition is caused by the lack of randomness in the corresponding measurement matrix $\A$.
The regular structures of (\ref{eq:matrix}) increase the risk that $\A$ does not satisfy the RIP requirement and  consequently affect the robustness of the recovery algorithm.

\begin{figure}[!t]
\centering
\begin{minipage}{\linewidth}
\centering
\subfloat[]{\includegraphics[width=.5\linewidth]{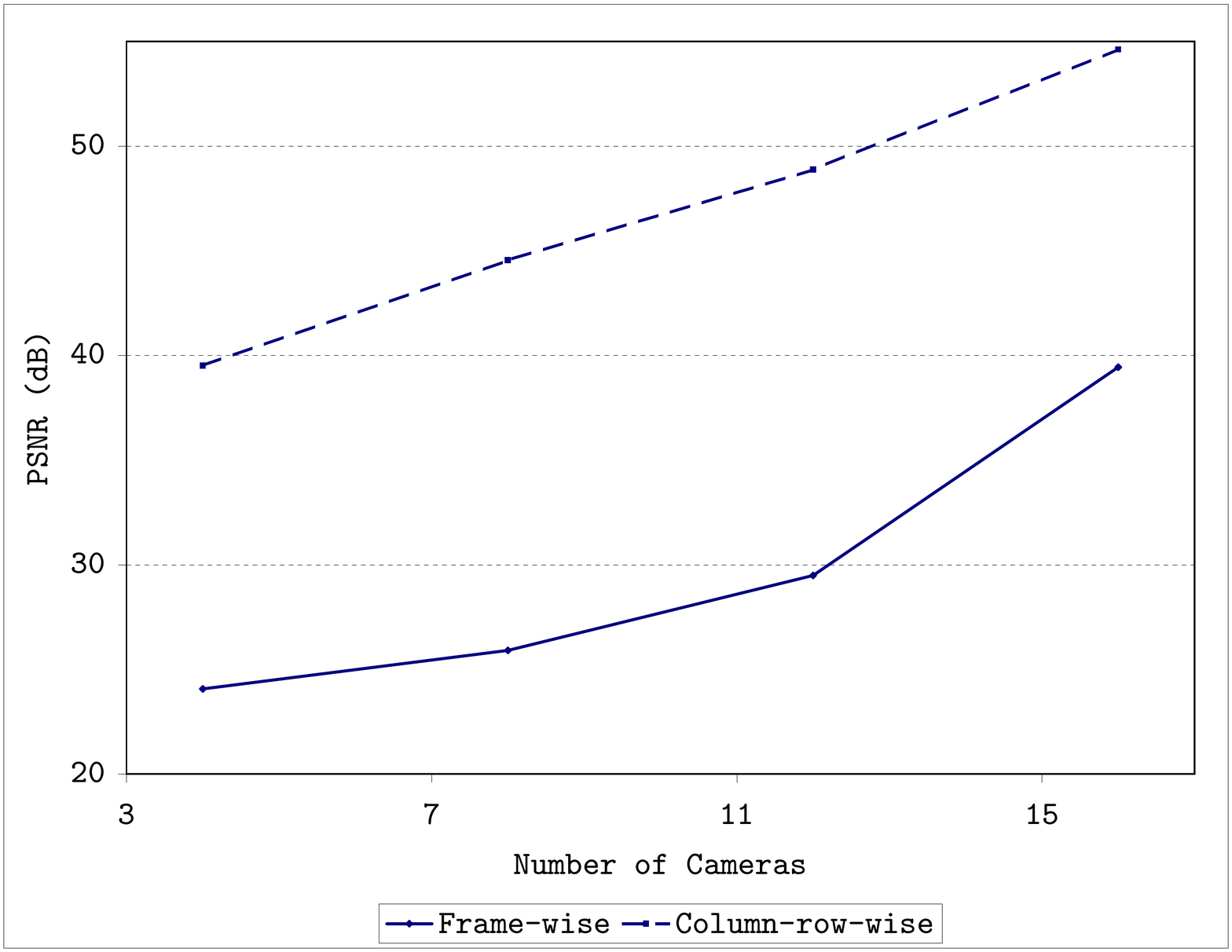}\label{fig:avg_psnr_cams}}
\hfil
\subfloat[]{\includegraphics[width=.5\linewidth]{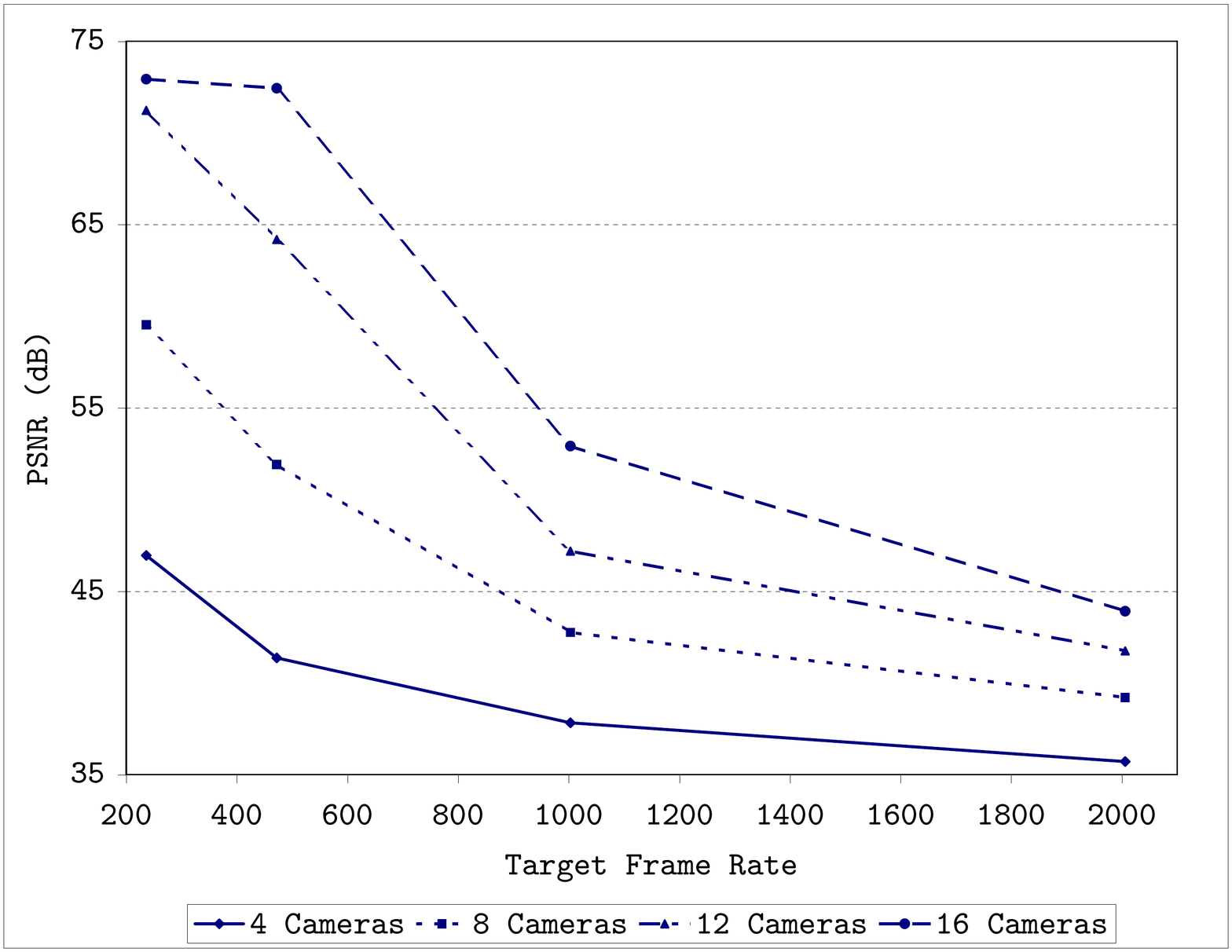}\label{fig:avg_psnr_frate}}
\end{minipage}
\caption{(a) Comparison of average PSNR vs. the number of cameras for ``Airbag" test video. (b) Comparison of average PSNR vs.\ target frame rate of column-row-wise coded acquisition for ``Bulb Bursting 1" test video using different number of cameras. The frame rate of the cameras is 60 frames/second.}
\label{fig:avg_psnr_cams_frate}
\end{figure}

In Fig.~\ref{fig:psnr:fl_cl_all}, the PSNR curves of
the column-row-wise coded acquisition are almost shifted versions of each other, suggesting a linear relation between the number of cameras and the PSNR of recovered HFV frames.
Fig.~\ref{fig:avg_psnr_cams} presents the average PSNR curves vs.\ the number of cameras for different coded acquisition techniques for a high-speed test video.  Similar patterns are found for all other test video signals.

To evaluate the performance of column-row-wise coded acquisition in relation to different target frame rates, we plot in Fig.~\ref{fig:avg_psnr_frate} the average PSNR as a function of the target frame rate for different number of cameras in the system.



\begin{figure*}[t]
\centering
\begin{minipage}{\linewidth}
\centering
\subfloat[Original]{\includegraphics[trim=207 37 187 54, clip, width=1.7in]{org_airbag1000fps_240x544_1003T59_sh1_f14}}%
\end{minipage}
\begin{minipage}{\linewidth}
\centering
\subfloat[Frame-wise with $4$ cameras]{\includegraphics[trim=207 37 187 54, clip, width=1.7in]{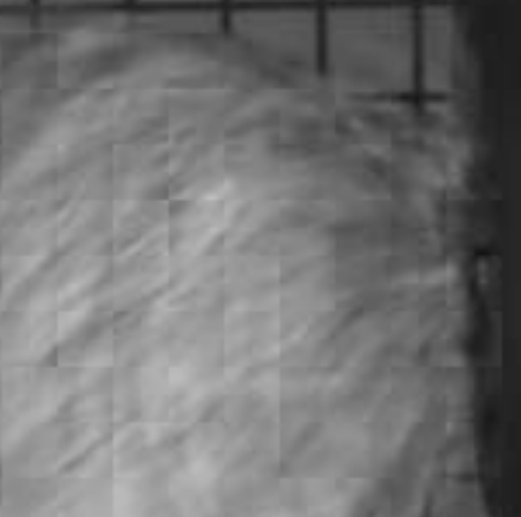}}%
\hfil
\subfloat[Pixel-wise with $4$ cameras]{\includegraphics[trim=207 37 187 54, clip, width=1.7in]{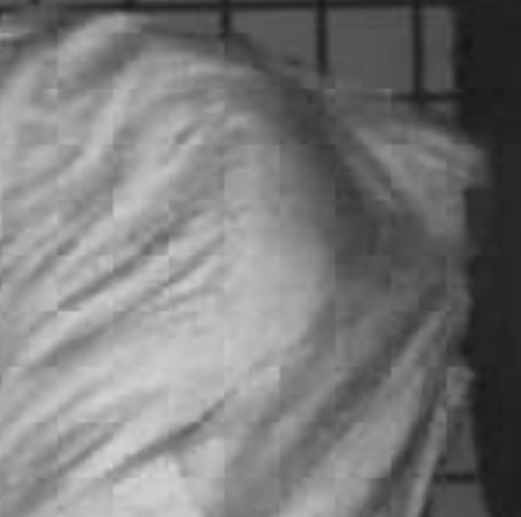}}%
\hfil
\subfloat[Column-row-wise with $4$ cameras]{\includegraphics[trim=207 37 187 54, clip, width=1.7in]{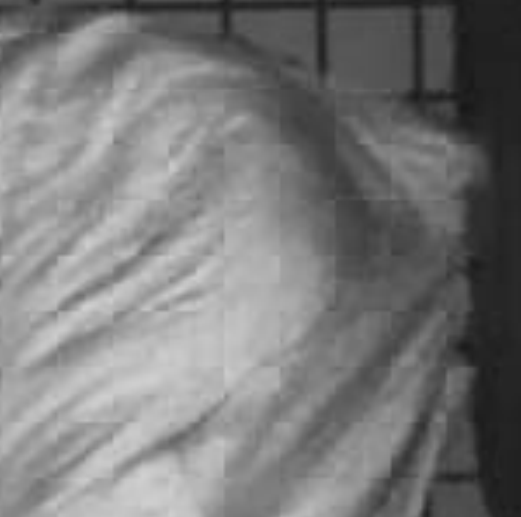}}%
\end{minipage}
\begin{minipage}{\linewidth}
\centering
\subfloat[Frame-wise with $8$ cameras]{\includegraphics[trim=207 37 187 54, clip, width=1.7in]{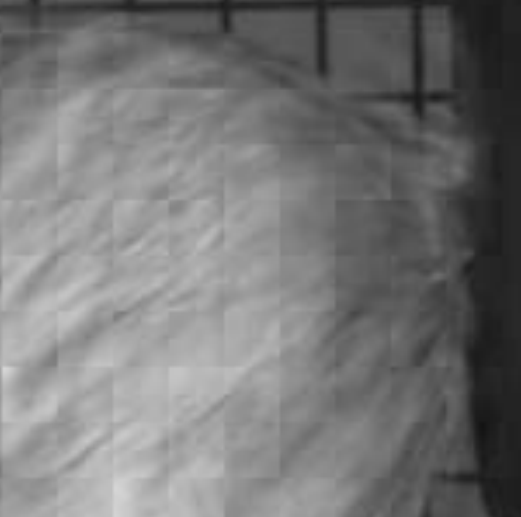}}%
\hfil
\subfloat[Pixel-wise with $8$ cameras]{\includegraphics[trim=207 37 187 54, clip, width=1.7in]{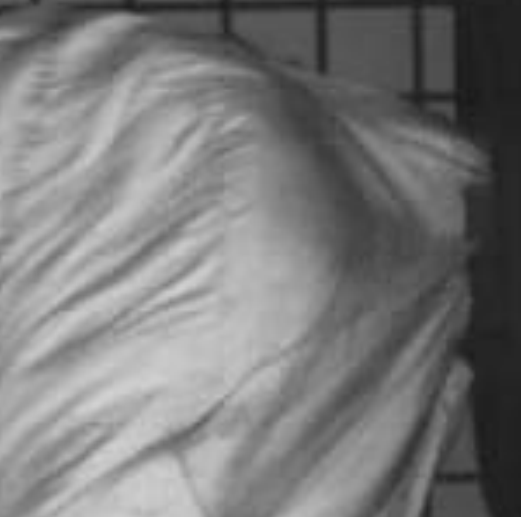}}%
\hfil
\subfloat[Column-row-wise with $8$ cameras]{\includegraphics[trim=207 37 187 54, clip, width=1.7in]{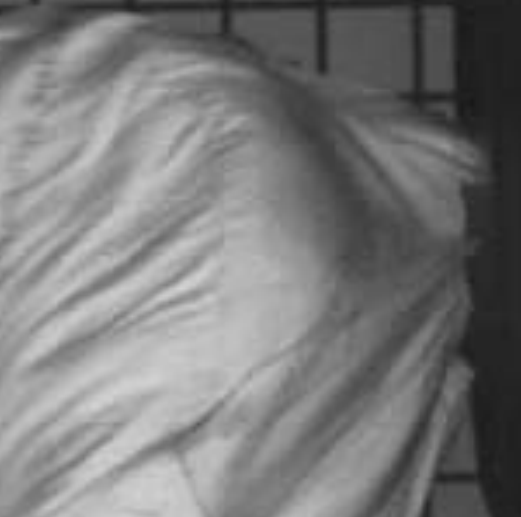}}%
\end{minipage}
\begin{minipage}{\linewidth}
\centering
\subfloat[Frame-wise with $16$ cameras]{\includegraphics[trim=207 37 187 54, clip, width=1.7in]{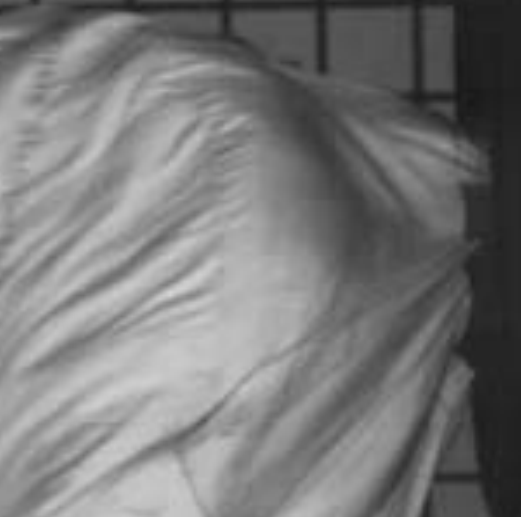}}%
\hfil
\subfloat[Pixel-wise with $16$ cameras]{\includegraphics[trim=207 37 187 54, clip, width=1.7in]{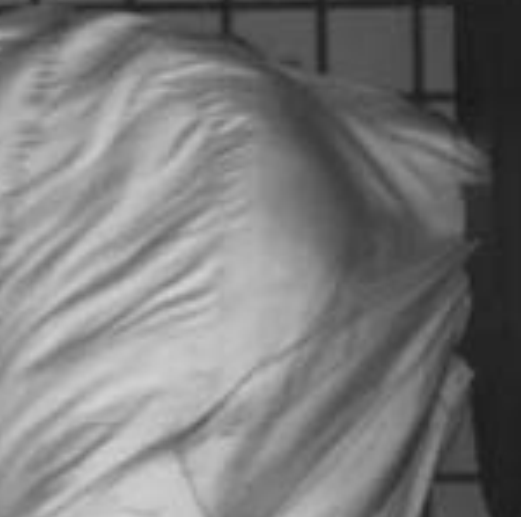}}%
\hfil
\subfloat[Column-row-wise with $16$ cameras]{\includegraphics[trim=207 37 187 54, clip, width=1.7in]{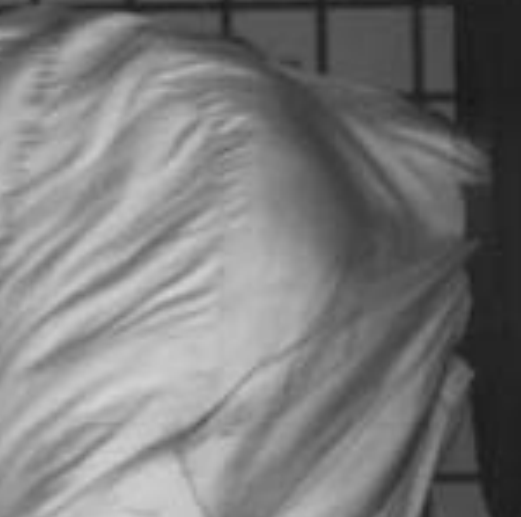}}%
\end{minipage}
\caption{Snapshots of recovered HFV sequence ``Airbag".}
\label{fig:airbag_snap}
\end{figure*}

\begin{figure*}[t]
\centering
\begin{minipage}{\linewidth}
\centering
\subfloat[Original]{\includegraphics[trim=62 8 268 82, clip, width=4.5cm]{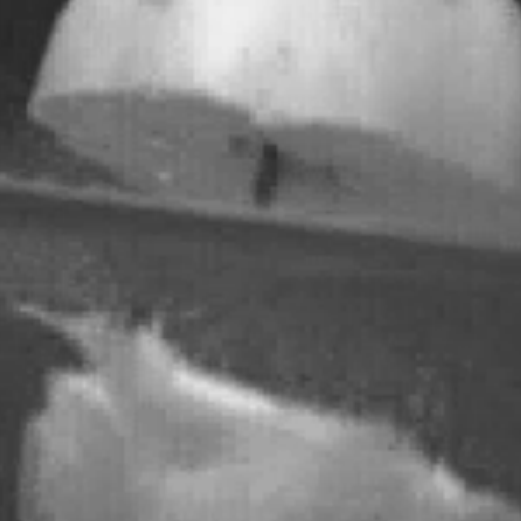}}
\end{minipage}
\begin{minipage}{\linewidth}
\centering
\subfloat[Frame-wise with $4$ cameras]{\includegraphics[trim=62 8 268 82, clip, width=4.5cm]{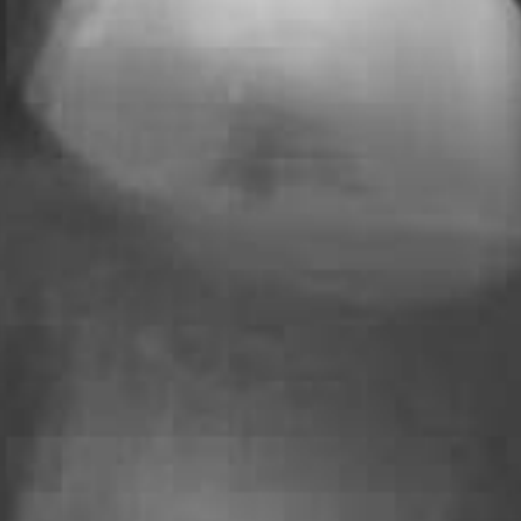}}%
\hfil
\subfloat[Pixel-wise with $4$ cameras]{\includegraphics[trim=62 8 268 82, clip, width=4.5cm]{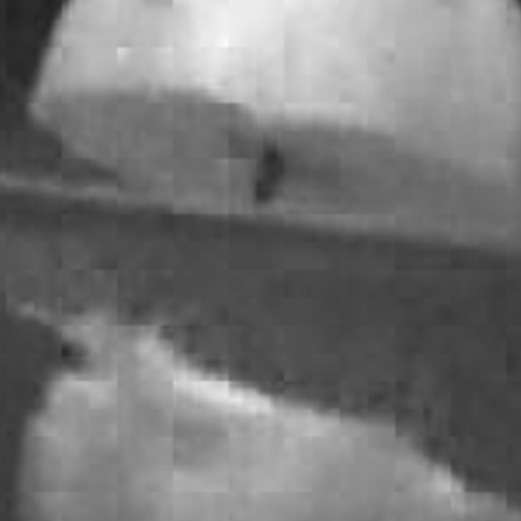}}%
\hfil
\subfloat[Column-row-wise with $4$ cameras]{\includegraphics[trim=62 8 268 82, clip, width=4.5cm]{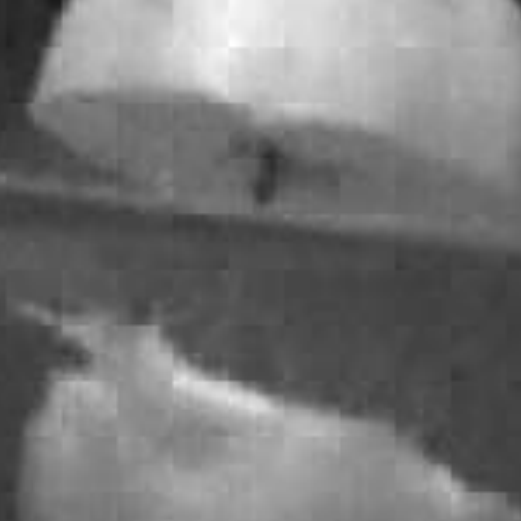}}%
\end{minipage}
\begin{minipage}{\linewidth}
\centering
\subfloat[Frame-wise with $8$ cameras]{\includegraphics[trim=62 8 268 82, clip, width=4.5cm]{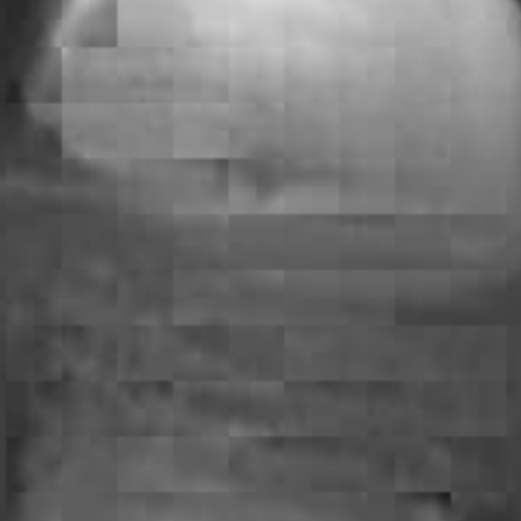}}%
\hfil
\subfloat[Pixel-wise with $8$ cameras]{\includegraphics[trim=62 8 268 82, clip, width=4.5cm]{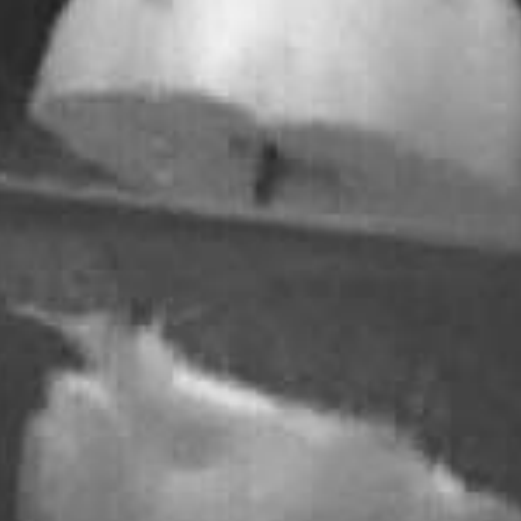}}%
\hfil
\subfloat[Column-row-wise with $8$ cameras]{\includegraphics[trim=62 8 268 82, clip, width=4.5cm]{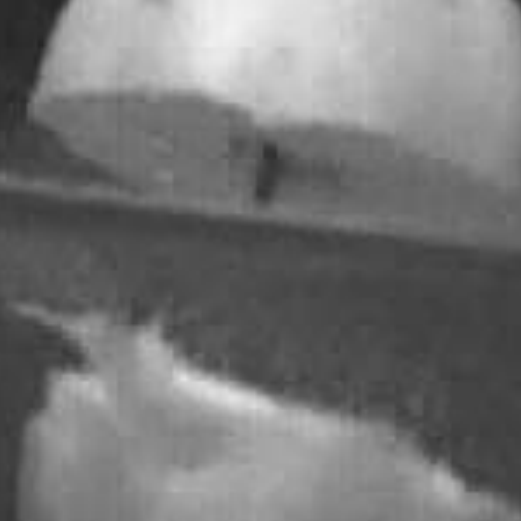}}%
\end{minipage}
\begin{minipage}{\linewidth}
\centering
\subfloat[Frame-wise with $16$ cameras]{\includegraphics[trim=62 8 268 82, clip, width=4.5cm]{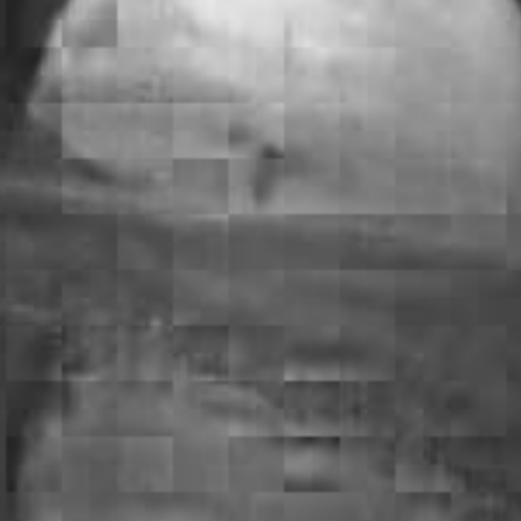}}%
\hfil
\subfloat[Pixel-wise with $16$ cameras]{\includegraphics[trim=62 8 268 82, clip, width=4.5cm]{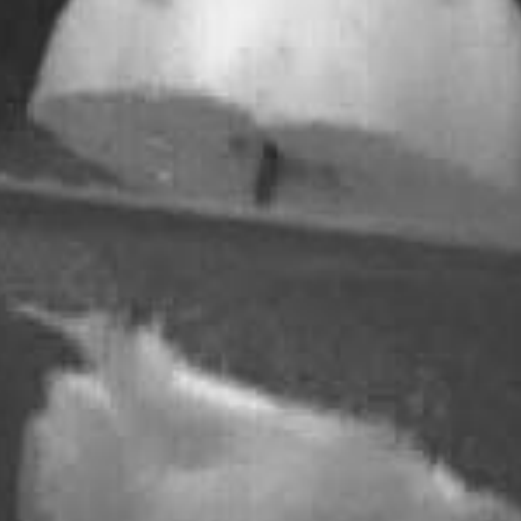}}%
\hfil
\subfloat[Column-row-wise with $16$ cameras]{\includegraphics[trim=62 8 268 82, clip, width=4.5cm]{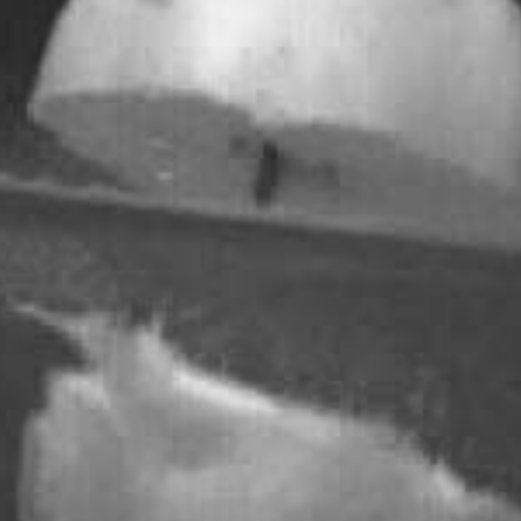}}%
\end{minipage}
\caption{Snapshots of recovered HFV sequence ``Apple Cutting".}
\label{fig:apple_snap}
\end{figure*}

Finally, in Fig.~\ref{fig:airbag_snap} and Fig.~\ref{fig:apple_snap} we present some snapshots of the recovered HFV in comparison with the original one for two different high-speed test video sequences.  The HFV snapshots are reproduced by $4$, $8$ and $16$ cameras and using frame-wise, pixel-wise, column-row-wise coded acquisition techniques.  As indicated by their virtually same PSNR performances,
the column-row-wise and pixel-wise acquisitions obtain indistinguishable HFV frames, and they reproduce much sharper images of less artifacts than the frame-wise coded acquisition.  The performance gap can be large when the number of cameras in the HFV acquisition system is small.

%
%

%


\section{Conclusions}
\label{sec:conclude}
We have studied a new methodology of acquiring high frame rate video using multiple
cameras of random coded exposure.  The objective of our research is to capture videos of both high temporal and spatial resolutions by inexpensive conventional cameras.  This is made possible by exploiting the sparsity of the video signal in time and space.  Three designs of multi-camera coded video acquisition with different cost-performance tradeoffs are discussed and their performance analyzed.  Simulation results are promising and they demonstrate the efficacy and potential of the new high-speed video acquisition methodology.

\appendix[Proof of the RIP for Measurement Matrix of Column-row-wise Coded Acquisition]

In this appendix we prove the RIP for random measurement matrix of column-row-wise coded acquisition (Theorem \ref{th:rip_DBD:Acr} in Section \ref{sec:pixel-col-row-wis}).
The proof here resembles that of Eftekhari {\it et al.} \cite{eftekhari2012ris} which applies a powerful theorem in \cite{krahmer2012suprema}.
First, we represent an special case of this theorem to facilitate our proof.

\begin{theorem}
Let $\mathcal{A}\subset\mathds{C}^{M\times N}$ be a set of matrices, and let $\cc$ be a Rademacher vector, whose entries are i.i.d. random variables that take the values $\pm1$ with equal probability. Denote by $\|.\|_F$ and $\|.\|_2$ Frobenious and spectral norms of a matrix. Set
\[
d_F(\mathcal{A}) = \sup_{\A\in\mathcal{A}}\|\A\|_F
\]
\[
d_2(\mathcal{A}) = \sup_{\A\in\mathcal{A}}\|\A\|_2
\]
and
\begin{eqnarray*}
E_1 &=& \gamma_2(\mathcal{A}, \|\cdot\|_2)\left(\gamma_2(\mathcal{A},\|\cdot\|_2)+d_F(\mathcal{A})\right)+d_F(\mathcal{A})d_2(\mathcal{A}) \\
E_2 &=& d_2(\mathcal{A})\left(\gamma_2(\mathcal{A},\|\cdot\|_2)+d_F(\mathcal{A})\right) \\
E_3 &=& d_2^2(\mathcal{A})
\end{eqnarray*}
Then, for $t>0$, it holds that
\begin{align*}
\Pr & \left(\sup_{\A\in\mathcal{A}}\left|\|\A\cc\|_2^2-\mathbb{E}\|\A\cc\|_2^2\right|\geq c_1E_1+t\right) \\
    &\quad\quad \leq 2\exp\left(-c_2\min\left(\frac{t^2}{E_2^2}, \frac{t}{E_3}\right)\right)
\end{align*}
where $c_1$ and $c_2$ are constants and $\mathbb{E}$ is the expectation of a random variable.
\label{th:base}
\end{theorem}
The term $\gamma_2(\mathcal{A}, \|\cdot\|_2)$ is $\gamma_2$-function of $\mathcal{A}$ which is a geometrical property of $\mathcal{A}$ and is widely used in the context of probability in Banach spaces \cite{ledoux1991probability}, \cite{talagrand2005generic}.

To prove RIP for random measurement matrix of column-row-wise coded acquisition we need to express the problem in the context of Theorem \ref{th:base}.
To do so, we need to change the entries of the random measurement matrix from $\{0, 1\}$ to $\{1, -1\}$.  This modification
can be done by adding another camera to capture the DC part of HFV signal (all entries in the measurement matrix of this camera are set to 1).
Subtracting the measurements taken by $K$ cameras from the DC part generates the new measurements made by the $\{-1, 1\}$ random matrices.
Also, we need to scale random measurement matrix by $1/\sqrt{K}$, making the elements of the random measurement matrix of each camera take on values in $\{-1/\sqrt{K}, 1/\sqrt{K}\}$.
In practice the scaling can be postponed until after random projection has been done, i.e., the scaling is applied to the measurements vector.

A few more definitions are needed in order to proceed with the proof. Let $N = TN_xN_y$ and $M = KN_xN_y$. For $\x \in \mathds{C}^N$, set $\f(\x) = \Psi\x$ where $\Psi$ is the Fourier transform matrix. Define $\Psi_{u,v} \in \mathds{C}^{T\times N}$ such that
\[
\Psi = \left[\Psi_{1,1}^* \cdots \Psi_{N_x,1}^* \Psi_{1,2}^* \cdots \Psi_{N_x,N_y}^*\right]^*
\]
where $\Psi_{u,v}^*$ is conjugate transpose of $\Psi_{u,v}$.
Also define $\f_{u,v}(\x) = \Psi_{u,v}\x \in \mathds{C}^T$ for $1\leq u\leq N_x$ and $1 \leq v \leq N_y$.
With these modifications, random measurement $y_k(u,v)$ is
\[
y_k(u,v) = \frac{1}{\sqrt{K}} \langle \rr_u^k\odot \cc_v^k, \f_{u,v}(\x) \rangle
\]
where vector $\rr_u^k$ ($\cc_v^k$) of size $T$ is the Rademacher sequence representing random coded exposure of row $u$ (column $v$) of camera $k$ in time, $\langle \cdot, \cdot \rangle$ and $\odot$ represent inner product and element-wise multiplication of two vectors respectively.

Now, by defining the set of all $S$-sparse signals with unit norm as
\[
\Omega_S = \left\{\x\in\mathds{C}^N:\|\x\|_0 \leq S, \|\x\|_2 = 1\right\}
\]
we can write the restricted isometry constant in Definition \ref{def:RIP_PSI} as
\[
\delta_S = \sup_{\x \in \Omega_S} \left|\|\A \f(\x)\|_2^2 -1\right|
\]
where we used the fact that with $\|\x\|_2 = 1$ we have
\[
\mathbb{E}\|\A\f(\x)\|_2^2  = 1.
\]

For $1\leq u \leq N_x$, $1\leq v\leq N_y$, $1\leq k\leq K$ define $\dot{\f}_{u,v}^k(\x) \in \mathds{C}^T$ as
\begin{equation*}
\dot{\f}_{u,v}^k(\x) = \R_u^k\f_{u,v}(\x) = \R_u^k\Psi_{u,v}\x
\end{equation*}
where $\R_u^k = diag\left(r_u^k(1), r_u^k(2), \cdots, r_u^k(T)\right)$ is a $T\times T$ random diagonal matrix representing random coded exposure of row $u$ of camera $k$ in time.
Also define $\dot{\F}_v^k(\x) \in \mathds{C}^{T\times N_x}$ as
\[
\dot{\F}_v^k(\x) = \left[\dot{\f}_{1,v}^k(\x) ,  \dot{\f}_{2,v}^k(\x) ,  \cdots , \dot{\f}_{N_x,v}^k(\x)\right].
\]
It can be easily verified that
\begin{equation}
\|\A\f(\x)\|_2^2 = \sum_{v,k} \|\dot{\F}_v^k(\x)^*\cdot\cc_v^k\|_2^2 = \|\dot{\F}(\x)\cdot \cc \|_2^2
\label{eq:norm2:Af}
\end{equation}
where $\cc = \left[{\cc_1^1}^*, \cdots, {\cc_1^K}^*, {\cc_2^1}^*, \cdots, {\cc_{N_y}^K}^*\right]^*$ is the random vector of length $TKN_y$ representing random coded exposure of columns of $K$ cameras and $\dot{F}(\x) = diag\left(\dot{\F}_1^1(\x)^*, \cdots, \dot{\F}_1^K(\x)^*, \cdots, \dot{\F}_{N_y}^K(\x)^*\right)$ is an $M\times KTN_y$ diagonal matrix.

In (\ref{eq:norm2:Af}), vector $\cc \in \mathds{R}^{TKN_x}$ is a Rademacher vector whose entries are i.i.d. random variables that take the values $\pm1$ with equal probability and the index set of the random process is $\mathcal{A} = \{\dot{F}(\x): \x \in \Omega_s\}$. Therefore the RIP of column-row-wise coded acquisition is now completely expressed in the settings of Theorem \ref{th:base}.

The next step is to estimate the quantities involved in Theorem \ref{th:base}. We start this part by estimating $\|\dot{\F}(\x)\|_2$:
\begin{eqnarray}
\|\dot{\F}(\x)\|_2 &=& \frac{1}{\sqrt{K}} \max_{v,k} \|\dot{\F}_v^k(\x)^*\|_2 \nonumber \\
&=&  \frac{1}{\sqrt{K}} \max_{v,k}\|\dot{\F}_v^k(\x)\|_2\label{eq:norm:dof_F:1} \\
&=& \frac{1}{\sqrt{K}} \max_{v,k} \left\|\sum_{i=1}^N x(i) \dot{\F}_v^k(e_i)\right\|_2 \label{eq:norm:dof_F:2}\\
&\leq& \frac{1}{\sqrt{K}} \max_{v,k} \sum_{i=1}^N|x(i)|\|\dot{\F}_v^k(e_i)\|_2 \label{eq:norm:dof_F:3}\\
&\leq& \frac{1}{\sqrt{K}} \max_{v,k} \left( \max_i\|\x\|_1\cdot\|\dot{\F}_v^k(e_i)\|_2\right) \label{eq:norm:dof_F:4} \\
&=&  \frac{1}{\sqrt{K}}\|\x\|_1\max_{i,v,k}\|\dot{\F}_u^k(e_i)\|_2 \label{eq:norm:dof_F:5}
\end{eqnarray}
where $\{e_i\}_{i=1}^N \in \mathds{R}^N$ are canonical basis. In passing from (\ref{eq:norm:dof_F:1}) to (\ref{eq:norm:dof_F:2}) we used the linearity of $\dot{\F}_v^k(.)$, passing from (\ref{eq:norm:dof_F:2}) to (\ref{eq:norm:dof_F:3}) follows from triangle inequality and (\ref{eq:norm:dof_F:4}) is the result of Holder inequality.

The next step is to estimate $\|\dot{\F}_v^k(e_i)\|_2$. Let $\dot{\F}_v^k(e_i)(t,u)$ be the entry of matrix $\dot{\F}_v^k(e_i)$ at row $t$ and column $u$. It is easy to verify that
\begin{align*}
\dot{\F}_v^k & (e_i)(t,u) \\
             & \quad = \frac{r_u^k(t)}{\sqrt{N}}\exp\left(-jd_i\left((v-1)N_xT+(u-1)T+t-1\right)\right) \\
             & \quad = \exp\left(-jd_i\left((v-1)N_xT-T-1\right)\right)\cdot \\
             & \quad\quad\quad\quad \frac{r_u^k(t)}{\sqrt{N}}\exp\left(-jd_it\right)\exp\left(-jd_iTu\right)
\end{align*}
where $j=\sqrt{-1}$ and $d_i = 2\pi(i-1)/N$. We can write matrix $\dot{\F}_v^k(e_i)$ as
\[
\dot{\F}_v^k(e_i) = \frac{\exp\left(-jd_i\left((v-1)N_xT-T-1\right)\right)}{\sqrt{N_xN_y}} \D_{l,i} \R^k \D_{r,i}
\]
where $\D_{l,i} = diag\left(\exp(-jd_i), \exp(-2jd_i), \cdots, \exp(-Tjd_i)\right)$ and $\D_{r,i} = diag\left(\exp(-Tjd_i), \cdots, \exp(-N_xTjd_i)\right)$ are $T\times T$ and $N_x\times N_x$ diagonal matrices and $\R^k \in \mathds{R}^{T\times N_x}$ is the random row modulation matrix
\[
\R^k = \frac{1}{\sqrt{T}}\left[\rr_1^k, \rr_2^k, \cdots, \rr_{N_x}^k\right]
\]
with $\rr_u^k \in \mathds{R}^T$, $1\leq u \leq N_x$, being Rademacher sequence representing random temporal coded exposure of camera $k$ at row $u$.
Therefore we have
\begin{align*}
\|\dot{\F}_v^k & (e_i)\|_2 \\
               & \quad = \left\|\frac{\exp\left(-jd_i\left((v-1)N_xT-T-1\right)\right)}{\sqrt{N_xN_y}}\D_t\R^k\D_u\right\|_2 \\
               & \quad = \frac{1}{\sqrt{N_xN_y}}\|\R^k\|_2
\end{align*}
%
%
%
%
Using the above result in (\ref{eq:norm:dof_F:5}), we have
\[
\|\dot{\F}(\x)\|_2 \leq \frac{1}{\sqrt{M}}\|\x\|_1 \max_{k}\|\R^k\|_2
\]


To complete the proof we need to calculate $d_F(\mathcal{A})$, $d_2(\mathcal{A})$ and $\gamma_2(\mathcal{A}, \|\cdot\|_2)$.
For $d_F(\mathcal{A})$ we have
\begin{eqnarray}
d_F(\mathcal{A}) &=& \sup_{\dot{\F}(\x)\in\mathcal{A}} \|\dot{\F}(\x)\|_F \nonumber \\
 &=& \sup_{\x \in \Omega_S} \frac{1}{\sqrt{K}}\sqrt{\sum_{v,k}\|\dot{F}_v^k(\x)\|_F^2} \nonumber \\
 &=& \sup_{\x \in \Omega_S} \frac{1}{\sqrt{K}}\sqrt{\sum_{v,k}\sum_u\|\dot{f}_u,v^k(\x)\|_2^2} \nonumber \\
 &=& \sup_{\x \in \Omega_S} \frac{1}{\sqrt{K}}\sqrt{\sum_{v,k}\sum_u\left(\x^*\Psi_{u,v}^*(\R_u^k)^2\Psi_{u,v}\x\right)} \nonumber \\
 &=& \sup_{\x \in \Omega_S} \frac{1}{\sqrt{K}}\sqrt{\x^*\left(\sum_k\left(\sum_{u,v}\Psi_{i,v}^*\Psi_{u,v}\right)\right)\x} \nonumber \\
 &=& \sup_{\x \in \Omega_S} \frac{1}{\sqrt{K}}\sqrt{K\x^*\x} = 1 \label{eq:d_F}
\end{eqnarray}

For $d_2(\mathcal{A})$ we have
\begin{eqnarray}
d_2(\mathcal{A}) &=& \sup_{\dot{F}(\x)\in\mathcal{A}} \|\dot{\F}(\x)\|_2 \nonumber \\
 &\leq& \frac{2}{\sqrt{M}} \max_{k}\|\R^k\|_2 \sup_{\x \in \Omega_s} \|\x\|_1 \nonumber \\
 &\leq& \sqrt{\frac{S}{M}}\max_{k}\|\R^k\|_2 \label{eq:d_2:2}
\end{eqnarray}
where (\ref{eq:d_2:2}) follows because for $\x\in\Omega_s$ we have $\|\x\|_2 = 1$ and $\|\x\|_0 = S$.

The following upper bound can be calculated for $\gamma_2(\mathcal{A}, \|\cdot\|_2)$ using Lemma 6 in \cite{eftekhari2012ris}:
\begin{equation}
\gamma_2(\mathcal{A}, \|\cdot\|_2) \leq c \sqrt{\frac{S}{M}}\log S\log N
\label{eq:gamma}
\end{equation}
where $c$ is a constant.

Now, our goal is to calculate $\Pr(\delta_S > \delta)$ for a prescribed $0<\delta<1$. Assuming $M \geq \delta^{-2}S\log^2S \log^2N$ and using (\ref{eq:d_F}), (\ref{eq:d_2:2}) and (\ref{eq:gamma}) we can compute $E_1$, $E_2$ and $E_3$ in Theorem~\ref{th:base}:
\begin{eqnarray}
E_1 &=& \gamma_2(\mathcal{A}, \|\cdot\|_2)\left(\gamma_2(\mathcal{A},\|\cdot\|_2)+d_F(\mathcal{A})\right)+d_F(\mathcal{A})d_2(\mathcal{A}) \nonumber \\
 &\leq& c\sqrt{\frac{S}{M}}\log S\log N\left(c\sqrt{\frac{S}{M}}\log S\log N+1\right) \nonumber \\
 & & \;\;\;\;\; +\sqrt{\frac{S}{M}}\max_{k}\|\R^k\|_2 \nonumber \\
 &\leq& c\delta(c\delta+1)+\frac{\delta}{\log S\log N}\max_k\|\R^k\|_2 \nonumber \\
 &\leq& c_3\delta+\frac{\delta}{\log S\log N}\max_k\|\R^k\|_2 \label{eq:E1:last}
\end{eqnarray}
where $c$ and $c_3$ are constants,
For $E_2$ we have
\begin{eqnarray*}
E_2 &=& d_2(\mathcal{A})\left(\gamma_2(\mathcal{A},\|\cdot\|_2)+d_F(\mathcal{A})\right) \\
 &\leq& \sqrt{\frac{S}{M}}\max_k\|\R^k\|_2\left(c \sqrt{\frac{S}{M}}\log S\log N +1\right) \\
 &\leq& \frac{\delta}{\log S\log N}\max_k\|\R^k\|_2(c\delta+1) \\
 &\leq& \frac{c_4\delta}{\log S\log N}\max_{k}\|\R^k\|_2
\end{eqnarray*}
and finally
\begin{eqnarray*}
E_3 &=& d_2^2(\mathcal{A}) \\
 &\leq&\frac{S}{M}\max_k\|\R^k\|_2^2 \leq \frac{\delta^2}{\log^2S\log^2N}\max_k\|\R^k\|_2^2.
\end{eqnarray*}
As it can be seen from above equations, the upper bounds of $E_1$, $E_2$ and $E_3$ depend on
\[
R_* = \max_k\|\R^k\|_2
\]
which is a random variable depending on the row random modulations of the $K$ cameras.

With above estimations and using Theorem~\ref{th:base}, we can now proceed to calculate the probability of RIP constant of random measurement matrix of column-row-wise coded exposure as follows:
\begin{align}
\Pr & \left(\delta_S \geq c_1(c_3+1)\delta + t \right) \nonumber \\
    & \quad \leq \Pr\left(\delta_S \geq c_1 \left(c_3+\frac{R_*}{\log S\log N}\right)\delta+t, \right. \nonumber \\
    & \quad\quad\quad\quad  \left.\vphantom{\left(c_3+\frac{R_*}{\log S\log N}\right)} R_* \leq \log S \log N \right) \nonumber \\
    & \quad \leq \Pr\left(\delta_S \geq c_1\left(c_3+\frac{R_*}{\log S\log N}\right)\delta+t, \right. \nonumber \\
    & \quad\quad\quad\quad  \left.\vphantom{\left(c_3+\frac{R_*}{\log S\log N}\right)} R_* \leq 2\right) \label{eq:prob:delta:R}\\
    & \quad \leq \Pr\left(\delta_S \geq c_1E_1 + t, R_* \leq 2\right) \label{eq:prob:delta:E1}\\
    & \quad = \Pr\left(\delta_S \geq c_1E_1 + t | R_* \leq 2\right)\Pr\left(R_* \leq 2\right) \nonumber \\
    & \quad \leq \Pr\left(\delta_S \geq c_1E_1 + t | R_* \leq 2\right) \nonumber \\
    & \quad \leq 2\exp\left(-c_2\min\left(\frac{t^2}{E_2^2},\frac{t}{E_3}\right)\right) \label{eq:prob:delta:th_base}
\end{align}
where in (\ref{eq:prob:delta:R}) we used the fact that
\[
\left(c_3+\frac{R_*}{\log S\log N}\right)\delta \leq (c_3+1)\delta
\]
if $R_* \leq \log S\log N$.  Lines (\ref{eq:prob:delta:E1}) and (\ref{eq:prob:delta:th_base}) follow from (\ref{eq:E1:last}) and Theorem~\ref{th:base} respectively.  Since $R_* \leq 2$ we have
\begin{eqnarray*}
E_2 & \leq & \frac{2c_4\delta}{\log S \log N} \\
E_3 & \leq & \frac{4\delta^2}{\log^2S\log^2N}.
\end{eqnarray*}
Therefore,
\begin{align*}
\Pr & \left(\delta_S \geq c_1(c_3+1)\delta + t \right) \\
    & \quad \leq 2\exp\left(-c_2 \log^2S\log^2N \min\left(\frac{t^2}{4c_4^2\delta^2},\frac{t}{4\delta}\right)\right).
\end{align*}
Substituting $t=\delta$ gives
\[
\Pr\left(\delta_S \geq (c_1(c_3+1)+1)\delta \right) \leq 2\exp\left(-c_0 \log^2S\log^2N \right)
\]
where $c_0 = c_2 \min \left((2c_4)^{-2}, 2^{-2}\right)$.
By redefining $\delta$ to absorb constant $(c_1(c_3+1)+1)$ we finally conclude that
\begin{align*}
\Pr & \left(\delta_S = \sup_{\x\in\Omega_S}\left|\|\A\f(\x)\|_2^2-1\right| \geq \delta\right) \\
    & \quad \quad \leq 2\exp\left(-c_0 \log^2S\log^2N \right) \label{eq:Pr:A_CR}
\end{align*}
which concludes the proof.

\ifCLASSOPTIONcaptionsoff
  \newpage
\fi

\bibliographystyle{IEEEtran}
\bibliography{refs}

\end{document}